\documentclass[aps,pra,twocolumn,superscriptaddress,notitlepage,nofootinbib,longbibliography, colorlinks=true]{revtex4-2}
\usepackage[utf8]{inputenc}
\usepackage{mathrsfs,graphicx,amsfonts,amssymb,amsmath,dcolumn,bm,xspace}
\usepackage[dvipsnames]{xcolor} 

\usepackage[colorlinks=true, allcolors=teal]{hyperref}
\usepackage{cleveref}
\begin{document}
\title{Nonreciprocal transmission in hybrid atomic ensemble-optomechanical systems}

\author{E. Kongkui Berinyuy}
\email{emale.kongkui@facsciences-uy1.cm}
\affiliation{Department of Physics, Faculty of Science, University of Yaounde I, P.O.Box 812, Yaounde, Cameroon}

\author{Jia-Xin Peng}
\affiliation{School of Physics and Technology, Nantong University, Nantong, 226019, People’s Republic of China}

\author{P. Djorwé}
\affiliation{Department of Physics, Faculty of Science, 
University of Ngaoundere, P.O. Box 454, Ngaoundere, Cameroon}
\affiliation{Stellenbosch Institute for Advanced Study (STIAS), Wallenberg Research Centre at Stellenbosch University, Stellenbosch 7600, South Africa}

\author{Abdourahimi}
\affiliation{Department of Physics, Faculty of Science, University of Yaounde I, P.O.Box 812, Yaounde, Cameroon}

\author{A.-H. Abdel-Aty}
\affiliation{Department of Physics, College of Sciences, University of Bisha, Bisha 61922, Saudi Arabia}

\author{K.S. Nisar}
\affiliation{College of Science and Humanities in Al-Kharj, Prince Sattam Bin Abdulaziz University, Al-Kharj 11942, Saudi Arabia}

\author{S. G. Nana Engo}
\affiliation{Department of Physics, Faculty of Science, University of Yaounde I, P.O.Box 812, Yaounde, Cameroon}

\begin{abstract}
We investigate perfect optical nonreciprocal transmission in a hybrid optomechanical system that incorporates an atomic ensemble. By introducing complex coupling strengths between the atomic ensemble and a mechanical oscillator, nonreciprocity is induced through interference between distinct optical pathways. The nonreciprocal transmission is governed by the real and imaginary components of the coupling constants, along with the relative phase differences between the optomechanical couplings. Our analysis reveals that, with precise tuning of system parameters, such as coupling strengths, detuning, and phase differences, perfect nonreciprocity can be achieved. We derive the conditions necessary for optimal nonreciprocal transmission and demonstrate its dependence on the complex nature of the coupling. These findings offer valuable insights for the design of nonreciprocal optical devices, including isolators and circulators, with potential applications in quantum communication, signal processing, and photonics.
\end{abstract}

\maketitle

\section{Introduction} \label{sec:Intr}

The interaction between atomic systems and mechanical oscillators has garnered considerable attention because of its potential in probing quantum phenomena and enabling novel applications. When coupled through radiation pressure, the mechanical system interacts with an optical field, which constitutes a cavity optomechanical (COM) system. These systems provide a versatile platform for exploring light-matter interactions at the quantum level. In particular, atomic ensembles trapped in optical cavities, which are strongly coupled to the optical field and mechanical oscillators, have demonstrated significant effects, such as quantum correlations and entanglement \cite{Meiser2006,Agasti2024,Bhattacherjee2009,Paternostro2010,jin2021macroscopic,Rostand2024}. These hybrid systems hold considerable promise for applications in quantum information processing \cite{Blais_2020,Stannigel_2012}, mass sensing \cite{Djorwe_2019,Tchounda_2023,Djor2024}, fundamental investigations of quantum mechanics at macroscopic scales \cite{Metcalfe:2014jnz,Aspelmeyer2014}, and nonlinear dynamics \cite{Xu_2024,Djor2022}. 

A key functionality for quantum technologies is nonreciprocity, which facilitates unidirectional signal transmission and is pivotal in devices such as optical isolators and circulators. Such nonreciprocal devices are required to prevent undesired reflections and protect sensitive optical components, such as lasers, from backscatter-induced instabilities \cite{Koch2010,Metelmann2015}. The realization of nonreciprocal transmission requires breaking of time-reversal symmetry, a condition that has been achieved theoretically and experimentally through various mechanisms, including magnetic fields, optomechanical interactions, and engineered dissipative processes \cite{Soljacic:03,He2018,el2018non,Oezdemir2019,Mbokop2024}. 

Recent advances in cavity optomechanics have unveiled new pathways for achieving nonreciprocal optical behavior without the reliance on magnetic effects. Specifically, robust optomechanical coupling has been shown to induce nonreciprocal transparency and phase shifts in microring resonators \cite{article8}. Additionally, the Lindblad master equation provides a rigorous framework for describing the dynamics of open quantum systems. Account for both unitary evolution and the effects of the external environment. Non-Hermitian Hamiltonians, which incorporate complex coupling terms, have been widely used to model nonreciprocal transmission in optical and mechanical systems, utilizing the interaction between loss and interference \cite{rotter2009non,ashida2020non,el2018non}. 

In this work, we investigate optical nonreciprocity in a hybrid atomic ensemble-cavity optomechanical system. By adjusting the system parameters—particularly the complex coupling strength $J_3$, which models dissipative interactions between the atomic ensemble and the mechanical mode—perfect nonreciprocity can be achieved. Our methodology employs a non-Hermitian Hamiltonian to describe the system, with its imaginary component $J_3$ symbolizing energy dissipation, resulting in transmission of unidirectional signals. This is consistent with recent advances in non-Hermitian physics, where dissipation-induced nonreciprocity has been experimentally demonstrated in optical cavities and waveguides \cite{el2018non, Oezdemir2019}. 

The structure of this paper is as follows: In Section \ref{sec:II}, we introduce the theoretical model of the hybrid atomic ensemble-cavity optomechanical system and derive its Hamiltonian and equations of motion. In Section \ref{sec:III}, we analyze the conditions required for nonreciprocal transmission and study the effects of system parameters on transmission properties. Finally, we conclude with a summary of our findings and discuss potential applications in quantum information processing and photonics.

\section{Perfect optical nonreciprocity} \label{sec:II}

\subsection{Model} \label{sec:IIA}

\begin{figure}[htp!]
\centering
\includegraphics[width=\linewidth]{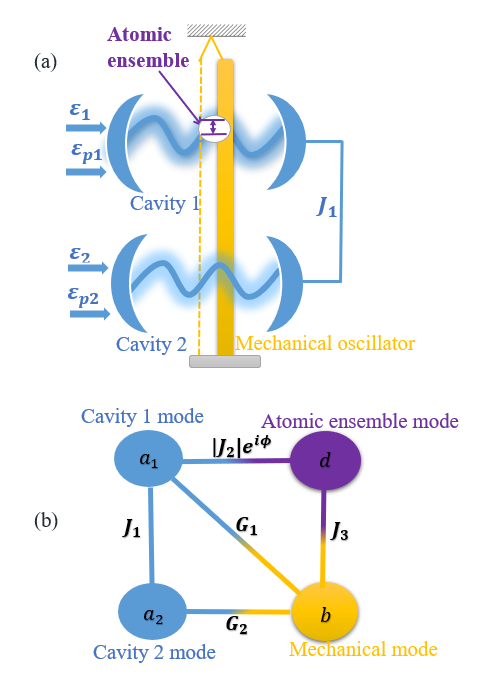}
\caption{(a) A hybrid optomechanical system with atomic ensemble in one cavity. The coupling fields (probe fields) with amplitudes $\mathcal{E}_1$ and $\mathcal{E}_2~(\mathcal{E}_{p1}$ and $\mathcal{E}_{p2})$ are used to drive the two cavities that are tunnel-coupled and coupled to the common mechanical mode via radiation pressure. (b) The diagram of interactions among subsystems in the optomechanical system.}
  \label{fig:setup}
\end{figure}

We consider a hybrid optomechanical system consisting of two coupled cavities, one of which contains a trapped atomic ensemble, as illustrated in Figure \ref{fig:setup}. The first cavity mode is coupled to the atomic ensemble, whereas the two cavities are connected via photon tunneling. A mechanical oscillator is coupled to both cavities through radiation pressure, enabling interactions between the optical field and mechanical motion. The system is driven by strong external fields and probed by weak fields to analyze its optical response.

The total Hamiltonian of the system include the following energy terms, the interactions among the subsystems, and the driving fields, and can be written as $(\hbar = 1)$,
\begin{equation}
H = H_o + H_{\rm en} + H_{\rm int} + H_{\rm driv}.
\end{equation}
The term $H_o$ stands for the free Hamiltonian of the cavity fields and mechanical oscillator, and reads,
\begin{equation}
H_o = \Delta_1 a_1^\dagger a_1 + \Delta_2 a_2^\dagger a_2 + \omega_m b^\dagger b,
\end{equation}
where $a_j (a_j^\dagger)$ and $b (b^\dagger)$ are the annihilation (creation) operators for the cavity and mechanical modes (with $j = 1, 2$), obeying the commutation relations $[a_j, a_j^\dagger] = 1$ and $[b, b^\dagger] = 1$. We assume that the number of atoms in the excited state is significantly lower than the total number of atoms $N$, indicating that the atomic ensemble operates under low excitation conditions. Consequently, we can treat the atomic ensemble as a bosonic mode. The atomic ensemble is initially modeled using spin-1/2 Pauli operators $\sigma_j^z$ and $\sigma_j^\pm$ for the $j^{th}$ two-level atom, which satisfy the commutation relations $[\sigma_j^+, \sigma_j^-] = \sigma_j^z$ and $[\sigma_j^z, \sigma_j^\pm] = \pm 2 \sigma_j^\pm$. To treat the atomic ensemble as a bosonic mode, we apply the Holstein-Primakoff transformation, which enables us to map the collective spin operators to bosonic creation and annihilation operators,
\begin{align*}
&\frac1N \sum_{j=1}^N \sigma_j^- = d, 
&& \frac1N \sum_{j=1}^N \sigma_j^+ = d^\dagger, 
&\sum_{j=1}^N \sigma_j^z = 2 d^\dagger d - N.
\end{align*}
The atomic ensemble interacts with both the optical and mechanical subsystems. Its Hamiltonian is written as:
\begin{equation}
\begin{aligned}\label{eq:3}
H_{\rm en} &= \frac{\Delta_{\rm en}}{2} \sum_{j=1}^N \sigma_j^z + \left( J_2^\ast a_1 \sum_{j=1}^N \sigma_j^+ + J_2 a_1^\dagger \sum_{j=1}^N \sigma_j^- \right) \\
&+ J_m \left( \sum_{j=1}^N \sigma_j^+ + \sum_{j=1}^N \sigma_j^- \right) (b + b^\dagger),
\end{aligned}
\end{equation}
where $J_2$ is the coupling strength between the atomic ensemble and the first cavity mode, and $J_m$ describes the coupling between the atomic ensemble and the mechanical oscillator. The detuning of the atomic ensemble mode from the driving frequency is defined as $\Delta_{\rm en} = \omega_{\rm en} - \omega_L$. 
The coupling $J_m$ is usually considered to be real in most quantum physics context. However, in systems that interact with an external environment such as bath, the coupling can become complex due to non-hermitian dynamics which include dissipation (loss) or amplification (gain). The system under consideration is model as an open quantum system where loss or gain is induced. The Hamiltonian of this type of system can acquire complex eigenvalues which represent the system that is not in equilibrium. The complex eigenvalues indicate that the Hamiltonian is non-hermitian, i.e., $H_{\rm com}\not=H_{\rm com}^\dagger$. Non-hermitian Hamiltonian usually have complex eigenvalues, where the imaginary part relates system's dissipation or amplification. The dynamics of the density matrix $\rho$ for an open quantum system is described by the Lindblad master equation, 
\begin{equation}
\dot{\rho}=-i[H_{\rm com},\rho]+\mathcal{D}[\rho],
\end{equation}	
where $\mathcal{D}[\rho]$ is the dissipator term that account for the interaction with the environment, 
\begin{equation}
\mathcal{D}[\rho]=\sum_{j=1}\Gamma_j\left(L_j\rho L_j^\dagger-\frac{1}{2}\{L_j^\dagger L_j,\rho\}\right).
\end{equation}	
In this work, our mechanical oscillator can lose energy to the environment or gain energy from the external laser field since it is interacting with the cavity. Because of the gain and loss induce in the system, $J_m$ can be treated as if had imaginary part to qualitatively describe these processes. In this light, $J_m$ takes the form $J_m=\text{Re}[J_m]+i\text{Im}[J_m]$. The Hamiltonian defining the interaction between atomic ensemble and mechanical oscillator takes the form 
\begin{equation}
H_{\rm com}=\left(\text{Re}[J_m]+i\text{Im}[J_m]\right)\left(\sum_{j=1}^N\sigma^+_j +\sum_{j=1}^N\sigma^-_j\right)(b+b^\dagger),
\end{equation}	
from the commutator relation, 	
\begin{equation}
\begin{split}
-i[H_{\rm com},\rho]=&-i\Big[\left(\text{Re}[J_m]+i\text{Im}[J_m]\right)\Big. \\
&\Big. \left(\sum_{j=1}^N\sigma^+_j+\sum_{j=1}^N\sigma^-_j\right)(b+b^\dagger),\rho\Big].
\end{split}
\end{equation}
Based on this expression, we have two contributions, one is from the real part and the other one is from the imaginary part respectively, represented as:
\begin{align}
&-i\left[\left(\text{Re}[J_3]\right)\left(\sum_{j=1}^N\sigma^+_j+\sum_{j=1}^N\sigma^-_j\right)(b+b^\dagger),\rho\right],\\
&-i\left[\left(i\text{Im}[J_3]\right)\left(\sum_{j=1}^N\sigma^+_j+\sum_{j=1}^N\sigma^-_j\right)(b+b^\dagger),\rho\right].
\end{align}
In the dissipative regime, that is, where the dissipative processes are dominant, we can neglect the real part of $J_m$ in order to emphasize on the effects of loss and gain. The last term in Eq. \eqref{eq:3} now becomes, 	
\begin{equation}
H_{com}=iJ_{m}\left(\sum_{j=1}^N\sigma^+_j+\sum_{j=1}^N\sigma^-_j\right)(b+b^\dagger).
\end{equation}	
For simplicity, we define $J_3=iJ_m$, this purely imaginary coupling renders the Hermitian Hamiltonian above non-Hermitian which is indicative of open quantum system exhibiting gain or loss. Therefore, we can consider the parameter $J_3$ as the degree of non-Hermiticity. The imaginary part of the coupling can introduce effective decay or amplification in the system's dynamics influencing the evolution of the density matrix through the dissipative term $\mathcal{D}[\rho]$. We note that the complex coupling $J_3$ is introduced as an effective parameter to capture the dissipative dynamics between the atomic ensemble and the mechanical oscillator in a non-Hermitian framework. This approach approximates the effects of environmental interactions, such as energy loss or gain, without explicitly modeling the bath dynamics through a full Lindblad master equation. In a complete Lindblad picture, the dissipative coupling would arise from the interplay between the Hermitian interaction Hamiltonian and the Lindblad dissipator terms, which describe specific dissipative channels (e.g., atomic decay or mechanical damping). The imaginary part of $J_3$ qualitatively represents the asymmetric energy exchange induced by the bath, which can be engineered, for example, by tuning the decay rates of atomic transitions using external laser fields. While our non-Hermitian treatment simplifies the analysis and highlights the role of dissipation in inducing nonreciprocity, it does not account for stochastic quantum jumps or the detailed bath structure, which would be captured in a full Lindblad formalism. This approximation is justified for studying the qualitative features of nonreciprocal transmission, as supported by similar approaches in the literature (e.g., Ref.~\cite{Takata2022}).

Thus, the Hamiltonian for the atomic ensemble can be rewritten as:
\begin{equation}
H_{\rm en} = \Delta_{\rm en} d^\dagger d + \left( J_2^\ast a_1 d^\dagger + J_2 a_1^\dagger d \right) + J_3 (d + d^\dagger)(b + b^\dagger).
\end{equation}
To provide insight into the microscopic origin of the complex coupling $J_3$, consider a scenario where the atomic ensemble and mechanical oscillator are coupled to a common bath, such as a thermal reservoir or an engineered environment. The Hermitian interaction between the atomic ensemble and the mechanical oscillator, typically of the form $H_{\text{\rm int}} = g (d + d^\dagger)(b + b^\dagger)$, is modified by the bath-induced dissipation. By adiabatically eliminating the bath degrees of freedom using the Born-Markov approximation, the effective interaction can acquire an imaginary component due to correlated dissipation or phase-dependent interference between the coherent coupling and dissipative channels. For instance, external laser fields can be used to control the atomic decay rates, introducing an effective dissipative coupling that renders the Hamiltonian non-Hermitian. This mechanism is analogous to the imaginary coupling between optical resonators induced by gain or loss, as demonstrated by \textcite{Takata2022}. Our phenomenological use of $J_3$ captures these effects in a simplified manner, enabling us to focus on the resulting nonreciprocal transmission without solving the full bath dynamics.

The interaction Hamiltonian between the cavity modes and the mechanical oscillator is given by,
\begin{equation}
H_{\rm int} = g_1 a_1^\dagger a_1 (b + b^\dagger) + g_2 a_2^\dagger a_2 (b + b^\dagger) + J_1 (a_1 a_2^\dagger + a_2 a_1^\dagger),
\end{equation}
where $g_j (j = 1, 2)$ are the optomechanical coupling strengths between the cavity modes and the mechanical oscillator, and $J_1$ represents the photon tunneling coupling between the two cavities. The driving fields are modeled by the Hamiltonian,
\begin{equation}
\begin{split}
H_{\rm driv} &= i(\mathcal{E}_1 a_1^\dagger - \mathcal{E}_1^\ast a_1) + i(\mathcal{E}_2 a_2^\dagger - \mathcal{E}_2^\ast a_2) \\
&+ i\mathcal{E}_{p1} (a_1^\dagger e^{-i\delta t} - a_1 e^{i\delta t}) + i\mathcal{E}_{p2} (a_2^\dagger e^{-i\delta t} - a_2 e^{i\delta t}),
\end{split}
\end{equation}
where $\mathcal{E}_1$ and $\mathcal{E}_2$ are the amplitudes of the driving fields for the cavities, $\mathcal{E}_{p1}$ and $\mathcal{E}_{p2}$ are the amplitudes of the probe fields, and $\delta$ is the detuning between the probe and driving fields.

The complex coupling strength $J_3$ plays a critical role in inducing nonreciprocal transmission. The real part of $J_3$ governs the interaction strength between the atomic ensemble and the mechanical oscillator, while the imaginary component represents dissipative processes, where energy is either lost to the environment or exchanged between subsystems. The introduction of dissipation through the imaginary part of $J_3$ breaks the symmetry of the system, leading to unidirectional signal transmission. The dissipation-induced nonreciprocity is a characteristic of non-Hermitian systems which can lead to nonreciprocal effects without the for traditionally used magnetic fields, although magnetic fields still play a vital role in certain systems to achieve nonreciprocal behaviour. A microscopic prescription involves coupling an atomic cloud for example, cold atoms or Bose-Einstein condensate to a mechanical oscillator such as a membrane using external laser fields. Dissipative mechanism is introduce into a cavity where mechanical oscillator and atomic ensemble are interacting via interaction with the bath (environment). By tuning the decay rate of the atomic transitions using external fields, one can introduce an effective dissipative coupling with an imaginary term. This renders the Hamiltonian non-hermitian and induces loss or gain into the system.

In Ref.~\cite{Takata2022}, the concept of imaginary coupling in non-Hermitian coupled-mode is explicitly demonstrated. The authors successfully studied the imaginary coupling between optical resonators induced by gain or loss. This study provides a theoretical framework for realizing complex couplings in quantum systems, which we extend to our optomechanical setup with an atomic ensemble. In our model, the imaginary coupling $J_3$ is introduced phenomenologically to capture similar dissipative effects, approximating the complex dynamics that would arise from a full Lindblad treatment of the system-bath interaction.
 
\subsection{Dynamical equations} \label{sec:IIB}

The dynamical evolution of the bosonic operators in the system is governed by the Heisenberg equations of motion, which are expressed as,
\begin{subequations}
\label{eq:Eq7}
\begin{align}
\dot{a_1} &= -(i\Delta_1 + \kappa_1 )a_1 - i g_1 a_1(b + b^\dagger) - iJ_1a_2 - iJ_2d \nonumber\\
& \quad+ \mathcal{E}_1 + \mathcal{E}_{p1} e^{-i\delta t},\label{eq:Eq7.a}\\
\dot{a_2} &= -(i\Delta_2 + \kappa_2 )a_2 - i g_2 a_2(b + b^\dagger) - iJ_1a_1 + \mathcal{E}_2 + \mathcal{E}_{p2} e^{-i\delta t},\label{eq:Eq7.b}\\
\dot{d} &= -(i\Delta_{\rm en} + f ) d - iJ^{\ast}_2a_1 - iJ_3(b + b^\dagger),\label{eq:Eq7.c}\\
\dot{b} &= -(i\omega_m + \gamma )b - i g_1 a_1^\dagger a_1 - i g_2 a_2^\dagger a_2 - i J_3(d + d^\dagger)
\label{eq:Eq7.d},
\end{align}
\end{subequations}
where the noise terms have been ignored. The first equation (Eq. \eqref{eq:Eq7.a}) describes the dynamics of the optical field in cavity 1, which is influenced by the mechanical oscillator through radiation pressure (coupling strength $g_1$) and by photon tunneling to cavity 2 (with coupling strength $J_1$). Additionally, the atomic ensemble couples to cavity 1 with strength $J_2$, and external driving fields $\mathcal{E}_1$ and $\mathcal{E}_{p1}$ act on the cavity. The terms $i \Delta_1 + \kappa_1$ account for the detuning and dissipation in cavity $1$. The second equation (Eq. \eqref{eq:Eq7.b}) stands for the optical field in cavity $2$, and it follows a similar form but with distinct coupling strengths and driving terms, reflecting its interaction with cavity $1$, the mechanical mode, and the external field $\mathcal{E}_2$. The atomic ensemble mode (Eq. \eqref{eq:Eq7.c}), represented by the operator $d$, evolves under the influence of the first cavity field $a_1$ (through coupling strength $J_2^*$) and the mechanical oscillator (via coupling strength $J_3$). The term $f$ models the dissipation within the atomic ensemble. The last equation (Eq. \eqref{eq:Eq7.d}) governs the dynamics of the mechanical oscillator. Its motion is influenced by both optical fields (through strengths $g_1$ and $g_2$) and by its coupling to the atomic ensemble through $J_3$. The terms $i \omega_m + \gamma$ account for the mechanical frequency and damping.

To analyze the system dynamics, we use the standard linearization technique. Under strong optical driving conditions, where the steady-state values dominate and fluctuations are small, we expand each Heisenberg operator as the sum of its steady-state value and a small fluctuation around it: 
\begin{align}
&a_j = \alpha_j + \delta a_j, && d = \rho + \delta d, & b = \beta + \delta b.
\end{align}
Substituting these expansions into Eq.~\eqref{eq:Eq7}, we obtain two sets of equations: one for the steady-state mean values and another for the fluctuation operators, which govern the quantum dynamics. The steady-state mean values $\alpha_j$, $\rho$, and $\beta$ are found by solving the following system of equations:
\begin{equation}
\begin{cases}
(i\Delta^\prime_1+\kappa_1)\alpha_1+iJ_1\alpha_2+iJ_2\rho-\mathcal{E}_1=0,\\
(i\Delta^\prime_2+\kappa_2)\alpha_2+iJ_1\alpha_1-\mathcal{E}_2=0,\\
(i\Delta_{\rm en}+f)\rho+iJ^\ast_2\alpha_1+2iJ_3\Re(\beta)=0,\\
(i\omega_m+\gamma)\beta+ig_1|\alpha_1|^2+ig_2|\alpha_2|^2+2iJ_3\Re(\rho)=0,
\end{cases}
\end{equation}
where $\Delta^\prime_j = \Delta_j + 2g_j \Re(\beta)$ are the effective detunings between the cavity modes and the coupling fields. In the steady-state regime, the mean values $\alpha_j$, $\rho$, and $\beta$ represent the amplitudes of the cavity fields, atomic ensemble, and mechanical displacement, respectively. The coupling terms between these subsystems introduce complex interactions: $J_2$ represents the interaction between the atomic ensemble and the cavity field, while $J_3$ models both conservative and dissipative interactions between the atomic ensemble and the mechanical oscillator. The steady-state equations indicate how energy is exchanged between these subsystems, with the imaginary part of $J_3$ driving dissipative energy exchange, which is critical for nonreciprocal behavior.

Next, we consider the quantum fluctuations around the steady state. The dynamics of the fluctuation operators are described by the linearized quantum Langevin equations, which take the form,
\begin{equation}
\label{Eq9}
\begin{cases}
\delta\dot{a_1}=-(i\Delta^\prime_1+\kappa_1)\delta a_1-i|J_2|e^{i\phi}\delta d-iJ_1\delta a_2\\-iG_1(\delta b+\delta b^\dagger)+\mathcal{E}_{p1}e^{-i\delta t},\\
\delta\dot{a_2}=-(i\Delta^\prime_2+\kappa_2)\delta a_2-iJ_1\delta a_1-iG_2e^{i\theta}(\delta b+\delta b^\dagger)\\+\mathcal{E}_{p2}e^{-i\delta t},\\
\delta\dot{d}=-(i\Delta_{\rm en}+f)\delta d-i|J_2|e^{-i\phi}\delta a_1-iJ_3(\delta b+\delta b^\dagger),\\
\delta\dot{b}=-(i\omega_m+\gamma)\delta b-iG_1(\delta a_1+\delta a^\dagger_1)-iG_2(e^{-i\theta}\delta a_2\\+e^{i\theta}\delta a^\dagger_2)-iJ_3(\delta d+\delta d^\dagger),
\end{cases}
\end{equation}
where the effective optomechanical coupling strengths are defined as $G_1 = g_1 \alpha_1$ and $G_2 = g_2 \alpha_2 e^{-i\theta}$. The phase difference between the effective optomechanical couplings is $\theta = \theta_2 - \theta_1$. Without loss of generality, we assume that $G_1$ is real, i.e., $\theta_1 = 0$, which reduces the phase difference to $\theta = \theta_2$. The quantum Langevin equations in Eq.\eqref{Eq9} describe the small fluctuations around the steady-state values and govern the quantum dynamics of the system. These linearized equations incorporate the coupling between the optical, mechanical, and atomic subsystems, as well as dissipation via the terms involving $\kappa_j$, $\gamma$, and $f$. The coupling terms involving $J_2$ and $J_3$ play a central role in generating nonreciprocity, with the dissipation modeled by the imaginary part of $J_3$ leading to asymmetric transmission. The full quantum dynamics, including these fluctuations, can be used to calculate observables such as the transmission coefficients, which reveal the nonreciprocal behavior of the system. These calculations are essential for understanding how energy dissipation and interference in the system give rise to unidirectional optical transmission. To solve Eq.~\eqref{Eq9}, we perform the transformations $\delta a_j \rightarrow \delta \tilde{a}_j e^{-i\Delta^\prime_j t}$, $\delta d \rightarrow \delta \tilde{d} e^{-i\Delta_{\rm en} t}$, and $\delta b \rightarrow \delta \tilde{b} e^{-i\omega_m t}$. These transformations yield,
\begin{equation}
\label{Eq10}
\begin{cases}
\delta\dot{\tilde{{a}}}_1=-\kappa_1\delta\tilde{a_1}-i|J_2|e^{i\phi}\delta\tilde{d}-iJ_1\delta\tilde{a_2}-iG_1\delta\tilde{b}+\mathcal{E}_{p1}e^{-iyt},\\
\delta\dot{\tilde{{a}}}_2=-\kappa_2\delta\tilde{a}_2-iJ_1\delta\tilde{a}_1-iG_2e^{i\theta}\delta\tilde{b}+\mathcal{E}_{p2}e^{-iyt},\\
\delta\dot{\tilde{d}}=-f\delta\tilde{d}-i|J_2|e^{-i\phi}\delta\tilde{a}_1-iJ_3\delta\tilde{b},\\
\delta\dot{\tilde{b}}=-\gamma\delta\tilde{b}-iG_1\delta\tilde{a}_1-iG_2e^{-i\theta}\delta\tilde{a}_2-iJ_3\delta\tilde{d}.
\end{cases}
\end{equation}
We now assume the solution to Eq.~\eqref{Eq10} takes the form,
\begin{equation}
\mathcal{O}=\mathcal{O}_+e^{-iy t}+\mathcal{O}_-e^{iy t}
\end{equation}
where $\mathcal{O} \equiv (\delta \tilde{a}_j, \delta \tilde{d}, \delta \tilde{b})$. Under the rotating wave approximation, where rapidly oscillating terms are neglected, we obtain a more compact matrix form as,
\begin{equation}
\mathcal{A}_1 \mathcal{X} = \mathcal{B},
\end{equation}
where,
\begin{equation}
\mathcal{A}_1=
\begin{pmatrix}
\kappa_1-iy&iJ_1&i|J_2|e^{i\phi}&iG_1\\
iJ_1&\kappa_2-iy&0&iG_2e^{i\theta}\\
i|J_2|e^{-i\phi}&0&f-iy&iJ_3\\
iG_1&iG_2e^{-i\theta}&iJ_3&\gamma-iy\\
\end{pmatrix},
\end{equation}
\begin{align}
&\mathcal{X}=
\begin{pmatrix}
\delta\tilde{a}_{1+}\\
\delta\tilde{a}_{2+}\\
\delta\tilde{d}_+\\
\delta\tilde{b}_+\\
\end{pmatrix},
&\mathcal{B}=
\begin{pmatrix}
\mathcal{E}_{p1}\\
\mathcal{E}_{p2}\\
0\\
0\\
\end{pmatrix}.
\end{align}
We can solve for $\mathcal{X}$ by inverting the matrix $\mathcal{A}_1$ as $\mathcal{X} = \mathcal{A}_1^{-1} \mathcal{B}$. This allows us to express the fluctuation components $\delta a_{1+}$ and $\delta a_{2+}$ as,
\begin{align}
&\delta a_{1+}=\frac{(i\tau_3+\tau_{4})\mathcal{E}_{p1}+(i\tau_1-\tau_2)\mathcal{E}_{p2}}{\mathcal{D}},\\
&\delta a_{2+}=\frac{(i\chi_1-\chi_2)\mathcal{E}_{p1}+(i\chi_3+\chi_{4})\mathcal{E}_{p2}}{\mathcal{D}},
\end{align}
where 
\begin{equation}
\begin{cases}
\tau_1=J_1y^2-J_1J^{2}_3-J_1\gamma f+G_1G_2ye^{-i\theta}+G_2|J_2|J_3e^{-i(\theta-\phi)}\\ \tau_2=J_1\gamma y+J_1 fy+G_1G_2fe^{-i\theta},\\ \tau_3=-G^2_2 y+y^3-J^2_3y-\gamma fy-\gamma y\kappa_2-fy\kappa_2,\\ \tau_{4}=G^2_2 f-\gamma y^2-fy^2-y^2\kappa_2+J^2_3\kappa_2+\gamma f\kappa_2,
\\
\chi_1=J_1y^2-J_1J^{2}_3-J_1\gamma f+G_1G_2ye^{i\theta}+G_2|J_2|J_3e^{i(\theta-\phi)},\\
\chi_2=J_1\gamma y+J_1 fy+G_1G_2fe^{i\theta},\\
\chi_3=-G^2_1 y+y^3-|J_2|^2 y-G_1|J_2|J_3e^{i\phi}-J^2_3y-\gamma f y\\-|J_2|G_1J_3e^{-i\phi}-\gamma y\kappa_1-fy\kappa_1,\\
\chi_{4}=-\gamma y^2-fy^2-y^2\kappa_1+J^2_3\kappa_1+\gamma f\kappa_1+|J_2|^2\gamma+G^2_1f,
\end{cases}
\end{equation}
and $\mathcal{D}$ is the determinant of $A_1$. Using the input-output formalism~\cite{XIA2019197}, the output field relations are given by:
\begin{equation}
\begin{aligned}
\mathcal{E}_{p1}^{out}+\mathcal{E}^{in}_{p1}e^{-iyt}=\sqrt{2\kappa_1}\delta\tilde{a}_1,\\
\mathcal{E}_{p2}^{out}+\mathcal{E}^{in}_{p2}e^{-iyt}=\sqrt{2\kappa_2}\delta\tilde{a}_2,
\end{aligned}
\end{equation}
where $\mathcal{E}_{p1,p2}^{in}=\mathcal{E}_{p1,p2}/\sqrt{2\kappa_j}$. For simplicity, we assume that the output fields also take the form: 
\begin{equation}
\mathcal{O}=\mathcal{O}_+e^{-iy t}+\mathcal{O}_-e^{iy t}.
\end{equation}
By comparing the coefficients of the terms $e^{-i y t}$ and $e^{i y t}$, we obtain the following relations:
\begin{equation}
\label{Eq22}
\begin{aligned}
&\mathcal{E}^{out}_{p1+}=\sqrt{2\kappa_1}\delta\tilde{a}_{1+}-\frac{\mathcal{E}_{p1}}{\sqrt{2\kappa_1}},
&\mathcal{E}_{p1-}^{out}=0,\\
&\mathcal{E}^{out}_{p2+}=\sqrt{2\kappa_2}\delta\tilde{a}_{2+}-\frac{\mathcal{E}_{p2}}{\sqrt{2\kappa_2}},
&\mathcal{E}_{p2-}^{out}=0.
\end{aligned}
\end{equation}
Particularly, perfect optical nonreciprocal transmission can be achieved if the transmission amplitudes satisfy the following conditions:
\begin{equation}
\label{Eq23}
\begin{aligned}
&\mathcal{T}_{12}=\left|\frac{\mathcal{E}_{p2+}^{out}}{\mathcal{E}_{p1}^{in}}\right|_{\mathcal{E}_{p2}^{in}=0}=1,
&\mathcal{T}_{21}=\left|\frac{\mathcal{E}_{p1+}^{out}}{\mathcal{E}_{p2}^{in}}\right|_{\mathcal{E}_{p1}^{in}=0}=0,\\
&\mathcal{T}_{12}=\left|\frac{\mathcal{E}_{p2+}^{out}}{\mathcal{E}_{p1}^{in}}\right|_{\mathcal{E}_{p2}^{in}=0}=0,
&\mathcal{T}_{21}=\left|\frac{\mathcal{E}_{p1+}^{out}}{\mathcal{E}_{p2}^{in}}\right|_{\mathcal{E}_{p1}^{in}=0}=1. 
\end{aligned}
\end{equation}
It is clear from these expressions that nonreciprocal transmission of optical signals can be achieved by adjusting the phase difference $\theta$ and the phase $\phi$ associated with the atoms. This will be explored in more detail in the next section.

\section{Results and discussion} \label{sec:III}

In this section, we analyze the conditions under which optical nonreciprocity can be achieved in the hybrid optomechanical system. We investigated the effects of key system parameters, such as coupling strengths, detuning, dissipation, and phase differences, on the transmission properties. The results reveal how tuning these parameters allows for control over nonreciprocal transmission, offering potential applications in optical isolation and signal routing.

\subsection{Optical nonreciprocal transmission}

\begin{figure}[htp!]
\centering
\includegraphics[width=.8\linewidth]{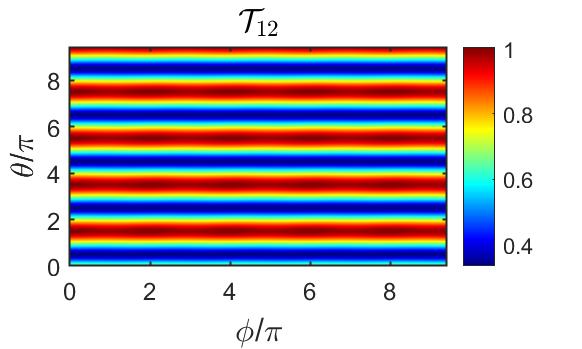}
\includegraphics[width=.8\linewidth]{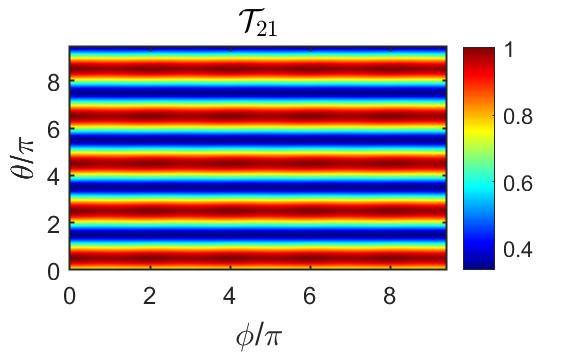}
\caption{Density plot of transmission amplitudes ($\mathcal{T}_{12}$ and $\mathcal{T}_{21}$) versus relative phase difference $\theta$ between effective optomechanical coupling strength and the phase $\phi$. We have used $G_1/\gamma=0.5$, $G_2=G_1$, $f/\gamma=10$, $\kappa_1/\gamma=2$, $\kappa_2=\kappa_1$, $J_1/\gamma=0.5$, $\omega_m/\gamma=10$, $\Delta^\prime_1=\Delta^\prime_2=\Delta_{\rm en}=\omega_m$, $|J_2|/\gamma=0.01$, and $J_3/\gamma=4.346e^{i\frac{\pi}{2}}$.}
  \label{fig:2}
 \end{figure}

The transmission amplitudes $\mathcal{T}_{12}$ and $\mathcal{T}_{21}$ are defined to delineate the transmission of optical signal from cavity 1 to cavity 2 and vice versa. These transmission coefficients are derived from the input-output relations and exhibit a strong dependence on the system parameters, particularly the phase differences between the effective optomechanical couplings and the coupling strengths of the atomic ensemble. \Cref{fig:2} displays the dependence of transmission amplitudes $\mathcal{T}_{12}$ and $\mathcal{T}_{21}$ on the phase difference $\theta$ between the effective optomechanical couplings and the phase $\phi$ associated with the atomic ensemble. In particular, although transmission amplitudes are not markedly influenced by phase $\phi$, both phases play a crucial role in the emergence of nonreciprocal transmission. 
The results in \Cref{fig:2} suggest that an optimal nonreciprocal transmission is achieved when the coupling constant between the atomic ensemble and the mechanical oscillator, $J_3$, is complex. The complex nature of $J_3$ introduces dissipation, which is essential to break the time-reversal symmetry and achieve nonreciprocity. Specifically, the imaginary component of $J_3$ encapsulates the energy exchange between the atomic ensemble and the mechanical oscillator, resulting in asymmetric transmission of optical signals.    

From a physical point of view, nonreciprocity arises from interference between two optical paths. One path is delineated along $a_1 \rightarrow d \rightarrow b \rightarrow a_2$, while the other follows $a_1 \rightarrow a_2$. Dissipative coupling $J_3$ enables unidirectional energy dissipation, thereby preventing backpropagation and facilitating nonreciprocal transmission.

\subsection{Influence of coupling strength and detuning}

\begin{figure*}[htp!]
\centering
\includegraphics[width=\textwidth]{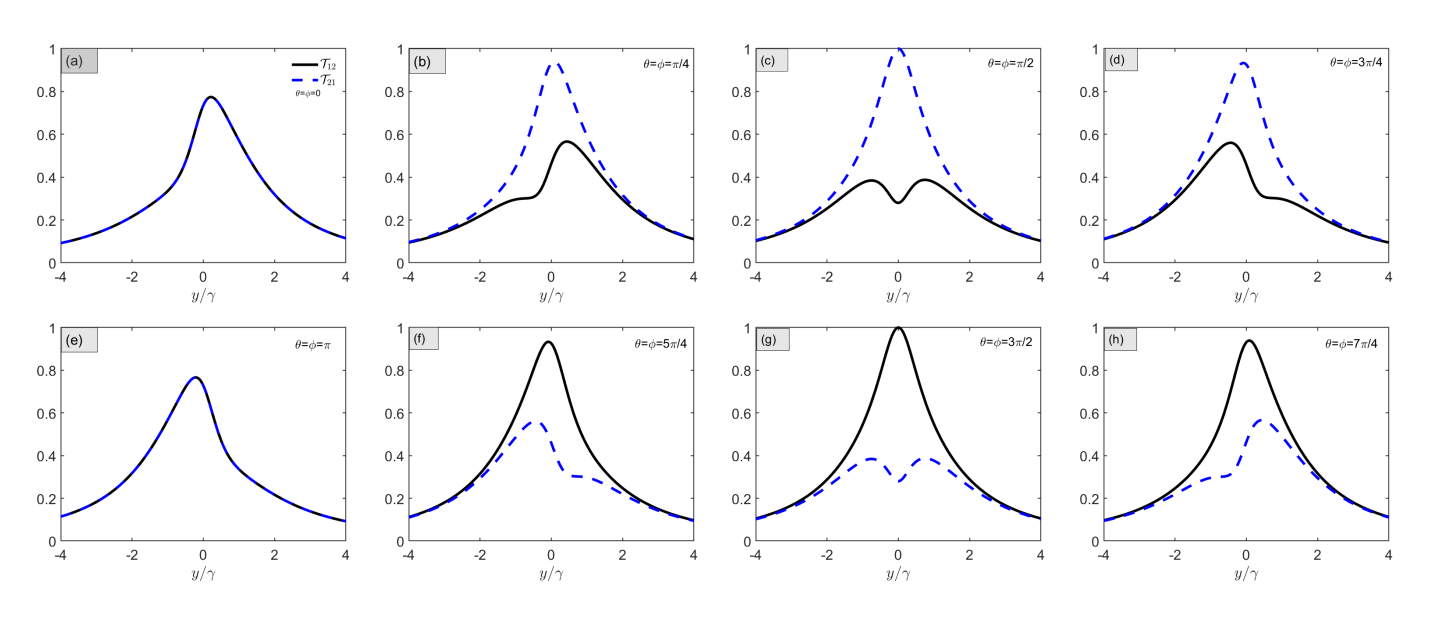}
\caption{Transmission amplitudes $\mathcal{T}_{12}$ and $\mathcal{T}_{21}$ as functions of detuning $y/\gamma$ for different effective phase differences $\theta$ and the phase $\phi$. (a) $\theta=\phi=0$, (b) $\theta=\phi=\frac{\pi}{4}$, (c) $\theta=\phi=\frac{\pi}{2}$, (d) $\theta=\phi=\frac{3\pi}{4}$, (e) $\theta=\phi=\pi$, (f) $\theta=\phi=\frac{5\pi}{4}$, (g) $\theta=\phi=\frac{3\pi}{2}$, and (h) $\theta=\phi=\frac{7\pi}{4}$. The other parameters are the same as in \Cref{fig:2}.}
\label{fig:3}
\end{figure*}

We now examine how the variation of the coupling strengths $J_2$ and $J_3$, as well as the detuning $\Delta_j$, affects the transmission properties of the system. The study presented in \Cref{fig:3} shows the behavior of transmission amplitudes $\mathcal{T}_{12}$ and $\mathcal{T}_{21}$ as functions of detuning $y = \delta - \omega_m$ for different values of phase difference $\theta$ and $\phi$. It is observed that when $\theta = \phi = m\pi$, where $m \in \mathbb{Z}$, the system exhibits reciprocal behavior, characterized by equal transmission amplitudes ($\mathcal{T}_{12} = \mathcal{T}_{21}$). This phenomenon occurs because the phase differences deplete the interference effect between the two optical paths. In contrast, when $\theta \neq m\pi$, the system manifests nonreciprocity with $\mathcal{T}_{12} \neq \mathcal{T}_{21}$. Optimal nonreciprocal transmission is achieved at resonance when $\theta = \phi = \frac{\pi}{2}$, where $\mathcal{T}_{12} \approx 0$ and $\mathcal{T}_{21} \approx 1$, or when $\theta = \phi = \frac{3\pi}{2}$, where $\mathcal{T}_{12} \approx 1$ and $\mathcal{T}_{21} \approx 0$. These findings illustrate that the direction of optical signal transmission can be regulated by adjusting the phase differences between subsystems. Specifically, for $0 < \theta < \pi$ and $0 < \phi < \pi$, the observation of $\mathcal{T}_{12} < \mathcal{T}_{21}$ implies stronger forward transmission from cavity 2 to cavity 1. Conversely, for $\pi < \theta < 2\pi$ and $\pi < \phi < 2\pi$, the reverse holds, with $\mathcal{T}_{12} > \mathcal{T}_{21}$ indicating enhanced transmission in the opposite direction.

\begin{figure*}[htp!]
\centering
\includegraphics[width=\textwidth]{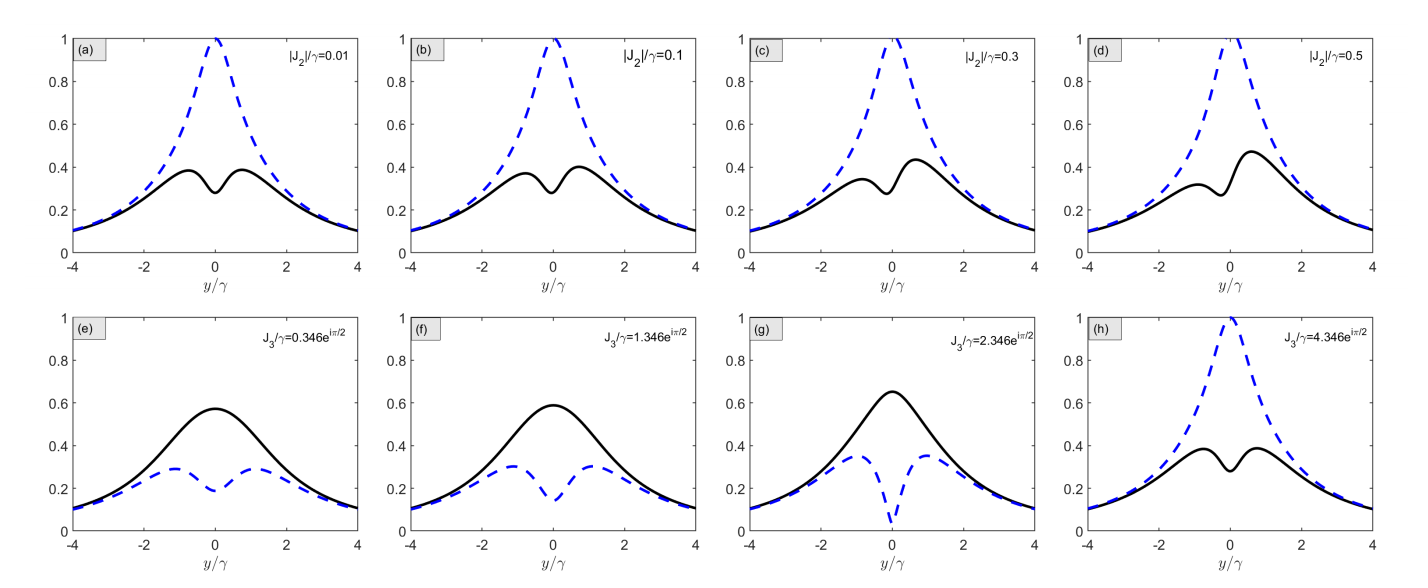}
\caption{Transmission amplitudes $\mathcal{T}_{12}$ and $\mathcal{T}_{21}$ as functions of detuning $y/\gamma$ for different coupling constants. (a) $|J_2|/\gamma=0.01$, (b) $|J_2|/\gamma=0.1$, (c) $|J_2|/\gamma=0.3$, (d) $|J_2|/\gamma=0.5$, (e) $J_3/\gamma=0.346e^{i\frac{\pi}{2}}$, (f) $J_3/\gamma=1.346e^{i\frac{\pi}{2}}$, (g) $J_3/\gamma=2.346e^{i\frac{\pi}{2}}$, and (h) $J_3/\gamma=4.346e^{i\frac{\pi}{2}}$. The other parameters are the same as in \Cref{fig:2}, except for $\theta=\phi=\frac{\pi}{2}$.}
  \label{fig:4}
\end{figure*}

Further investigation focuses on how variations in the coupling constants $|J_2|$ and $J_3$ affect the nonreciprocal response. As demonstrated in \Cref{fig:4}, increasing the coupling constant $|J_2|$, which regulates the interaction between the atomic ensemble and the first cavity mode, primarily affects the transmission amplitude $\mathcal{T}_{12}$. Alternatively, decreasing the coupling constant $J_3$ between the atomic ensemble and the mechanical mode results in a reverse transmission of the input signal, as illustrated in \Cref{fig:4} (gh). 

These results underscore the critical reliance of nonreciprocity on the coupling constants. When $J_3$ is adjusted to possess a significant imaginary component, dissipation assumes a dominant role, inhibiting backscattered signals and facilitating strong nonreciprocal transmission. In contrast, reducing dissipation (that is, decreasing $\Im(J_3)$) mitigates the nonreciprocal effect, prompting the system to display more reciprocal behavior.

\subsection{Conditions for perfect nonreciprocity}

We proceed to examine the conditions required to achieve perfect nonreciprocal transmission within the hybrid optomechanical system. The perfect nonreciprocity is realized when the transmission amplitude in one direction reaches unity ($\mathcal{T}_{12} = 1$), while in the reverse direction it is completely suppressed ($\mathcal{T}_{21} = 0$), or vice versa. This functionality is fundamental to applications in optical isolation and circulators within quantum information processing systems, where unidirectional light transmission is critical. 

Perfect nonreciprocal transmission can be realized through strategic adjustments of several parameters: (i) the phase difference $\theta$ between the effective optomechanical couplings; (ii) the coupling strength $J_3$ between the atomic ensemble and the mechanical oscillator, with particular attention to its complex nature, where the real part signifies conservative interactions and the imaginary part indicates dissipation; (iii) the relative coupling strength $|J_2|$, which facilitates the interaction between the atomic ensemble and cavity 1; and (iv) the dissipation rates of the mechanical oscillator and the atomic ensemble, represented by $\gamma$ and $f$, respectively. Induction of sufficient dissipation through the imaginary part of $J_3$ is essential to disrupt the symmetry of time-reversal. 

For simplicity, we consider the scenario where the phase differences are configured to $\theta = \phi = \frac{\pi}{2}$, a setup that generally produces pronounced nonreciprocal behavior. By defining $G_1 = \sqrt{\gamma \kappa_1}$ and $G_2 = \sqrt{\gamma \kappa_2}$, the necessary coupling strengths are determined by: 
\begin{equation}
	\begin{cases}
	J_1=\frac{G_1G_2}{\gamma+f},\\	
	 |J_2|=\frac{J_1J_3^2+J_1\gamma f-G_1G_2 f}{G_2 J_3},\\
	 J_3=\pm\sqrt{\frac{-R_8\pm\sqrt{R^2_8-4R_7R_9}}{2R_7}},
	\end{cases}
\end{equation}  
The expressions above are obtained with the help of conditions $\mathcal{T}_{12}=1$ and $\mathcal{T}_{21}=0$ (the expressions of $R_7, R_8,$ and $ R_9$ are presented in the appendix).

The dissipation introduced through the imaginary part of $J_3$ is instrumental in achieving perfect nonreciprocity. This dissipative interaction enables asymmetric energy exchange between the atomic ensemble and the mechanical oscillator, effectively mitigating back-reflections and facilitating unidirectional transmission. Conversely, reducing the dissipation (i.e., minimizing $\Im(J_3)$) diminishes the nonreciprocal effect and tends to reinstate reciprocal transmission, where $\mathcal{T}_{12} = \mathcal{T}_{21}$.

\begin{figure*}[htp!]
\centering
\includegraphics[width=\textwidth]{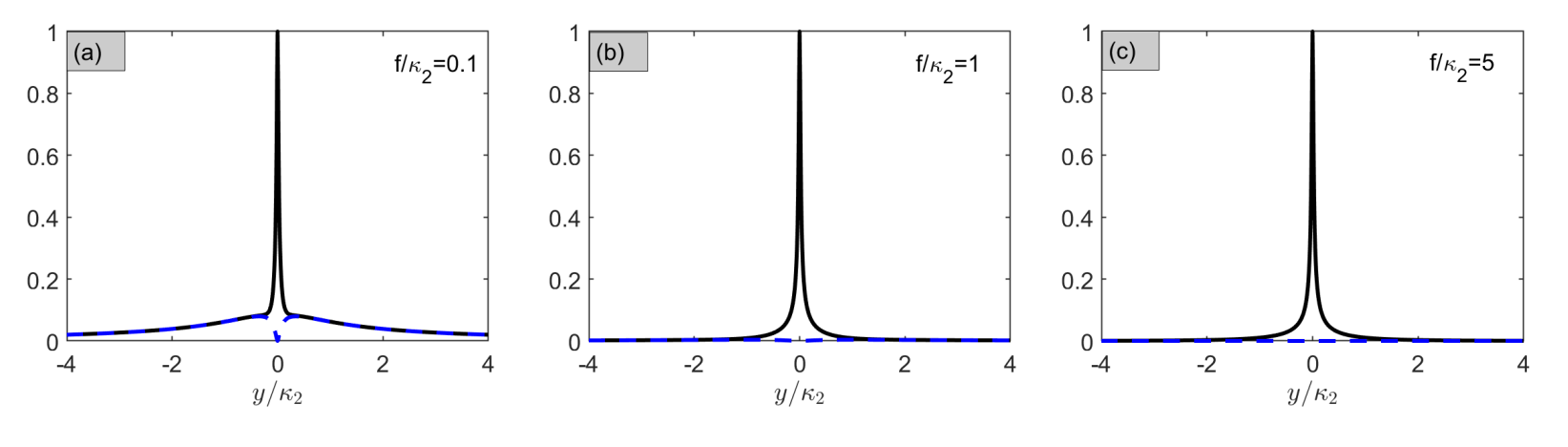}
\caption{Transmission amplitudes $\mathcal{T}_{12}$ and $\mathcal{T}_{21}$ as functions of $y/\kappa_2$ for different values of atomic ensemble decay rate $f$. (a) $f/\kappa_2=0.1$, (b) $f/\kappa_2=1$, (c) $f/\kappa_2=5$. The other parameters are the same as in \Cref{fig:2}, except for $\gamma/\kappa_2=0.01$, $\kappa_1/\kappa_2=1$, and $J_3=\sqrt{\frac{-R_8+\sqrt{R^2_8-4R_7R_9}}{2R_7}}$.}
  \label{fig:5}
\end{figure*}

\begin{figure*}[htp!]
\centering
\includegraphics[width=\textwidth]{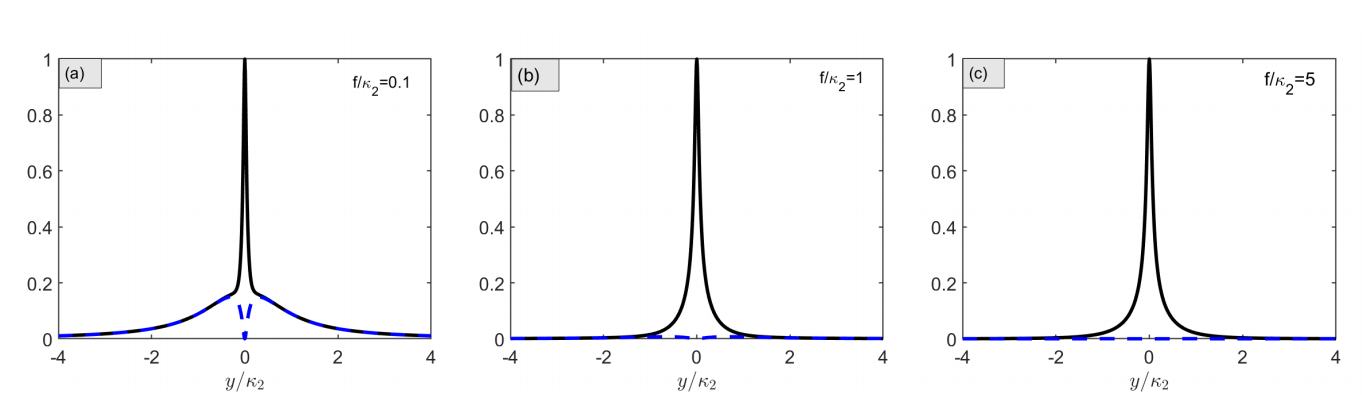}
\caption{Transmission amplitudes $\mathcal{T}_{12}$ and $\mathcal{T}_{21}$ as functions of $y/\kappa_2$ for different values of atomic ensemble decay rate $f$. (a) $f/\kappa_2=0.1$, (b) $f/\kappa_2=1$, (c) $f/\kappa_2=5$. The other parameters are the same as in \Cref{fig:2}, except for $\gamma/\kappa_2=0.01$, $\kappa_1/\kappa_2=1$, and $J_3=-\sqrt{\frac{-R_8-\sqrt{R^2_8-4R_7R_9}}{2R_7}}$.}
  \label{fig:6}
\end{figure*}

\subsection{Influence of atomic ensemble decay and mechanical damping}

To better understand the impact of system parameters on nonreciprocal transmission, this study investigates the influence of the decay rate of the atomic ensemble $f$ and the mechanical damping rate $\gamma$ on transmission amplitudes $\mathcal{T}_{12}$ and $\mathcal{T}_{21}$. These parameters dictate the dissipation within the atomic and mechanical subsystems and significantly affect the stability of nonreciprocity.  Figures \ref{fig:5} and \ref{fig:6} illustrate that perfect nonreciprocity is maintained throughout a spectrum of atomic decay rates $f$, indicating that minor fluctuations in the decay rate do not substantially alter the transmission characteristics. In fact, Figures \ref{fig:5} (a) and \ref{fig:6} (a) demonstrate that at reduced decay rates $f/\kappa_2 = 0.1$, the transmission amplitudes $\mathcal{T}_{12}$ and $\mathcal{T}_{21}$ show pronounced peaks near the resonance. However, nonreciprocal transmission is confined to a limited detuning range, suggesting heightened sensitivity to slight detuning variations when the atomic decay rate is minimal. With an increase in the decay rate of the atomic ensemble to $f/\kappa_2 = 1$, as depicted in Figures \ref{fig:5}(b) and \ref{fig:6}(b), the range of detuning allowing nonreciprocal transmission expands. This results in amplification of transmission $\mathcal{T}_{12}$ while $\mathcal{T}_{21}$ decreases markedly, indicating enhanced nonreciprocity. This expanded detuning range suggests that the system is more resilient to changes in detuning. For elevated atomic decay rates ($f/\kappa_2 = 5$), as shown in Figures \ref{fig:5}(c) and \ref{fig:6}(c), the system approaches near-perfect nonreciprocity over a broader detuning range. The peak magnitudes of $\mathcal{T}_{12}$ and $\mathcal{T}_{21}$ become less pronounced, implying that the system can maintain nonreciprocity. This infers that dissipation introduced by the atomic ensemble contributes to a stabilizing effect in preserving nonreciprocal transmission.  The findings of Figures \ref{fig:5} and \ref{fig:6} reveal that the higher decay rates of the atomic ensemble $f$ produce wider nonreciprocal transmission bands, thus strengthening the system's resilience to detuning. This is consistent with the idea that increased dissipation helps suppressing back-reflections and fosters stronger nonreciprocity.

\begin{figure*}[htp!]
\centering
\includegraphics[width=\textwidth]{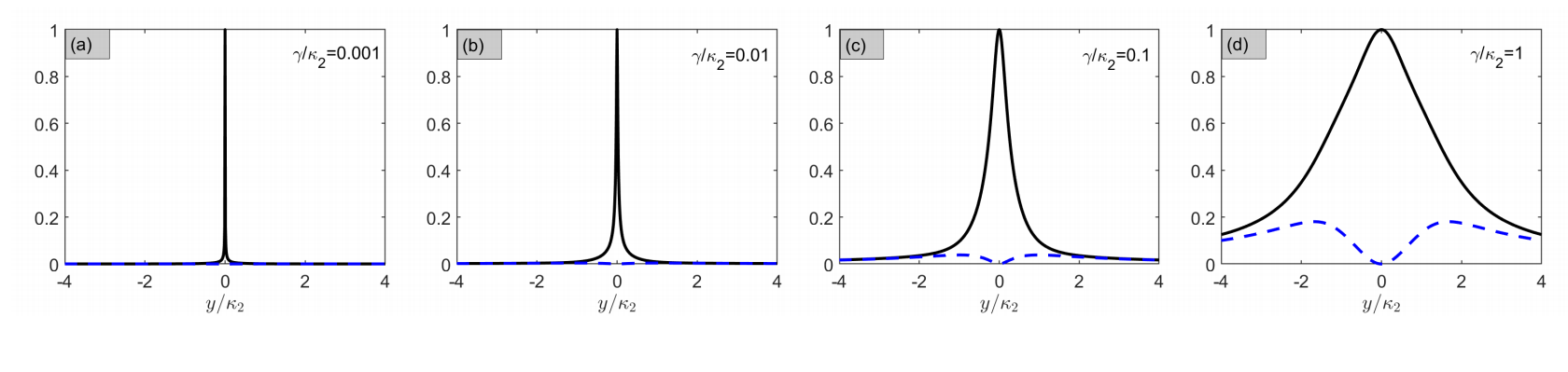}
\caption{Transmission amplitudes $\mathcal{T}_{12}$ and $\mathcal{T}_{21}$ as functions of detuning y$/\kappa_2$ for different values of mechanical damping rate $\gamma$. (a) $\gamma/\kappa_2=0.001$, (b) $\gamma/\kappa_2=0.01$, (c) $\gamma/\kappa_2=0.1$, (d) $\gamma/\kappa_2=1$. The other parameters are the same as in \Cref{fig:2}, except for $f/\kappa_2=1$, $J_3=\sqrt{\frac{-R_8+\sqrt{R^2_8-4R_7R_9}}{2R_7}}$ and $\kappa_1/\kappa_2=1$.}
  \label{fig:7} 
\end{figure*}

\begin{figure*}[htp!]
\centering
\includegraphics[width=\textwidth]{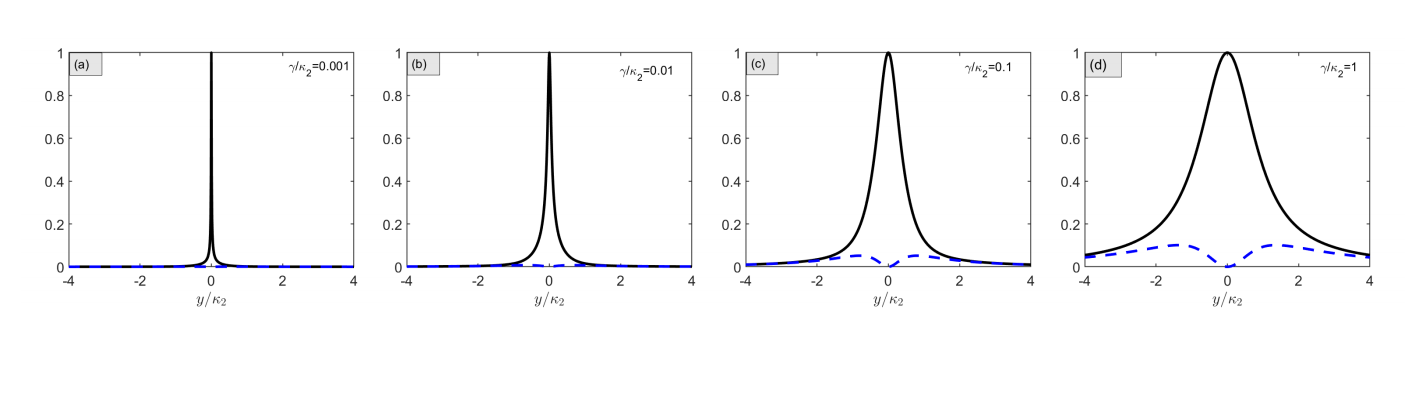}
\caption{Transmission amplitudes $\mathcal{T}_{12}$ and $\mathcal{T}_{21}$ as functions of detuning y$/\kappa_2$ for different values of mechanical damping rate $\gamma$. (a) $\gamma/\kappa_2=0.001$, (b) $\gamma/\kappa_2=0.01$, (c) $\gamma/\kappa_2=0.1$, (d) $\gamma/\kappa_2=1$. The other parameters are the same as in \Cref{fig:2} except for $f/\kappa_2=1$, $J_3=-\sqrt{\frac{-R_8-\sqrt{R^2_8-4R_7R_9}}{2R_7}}$ and $\kappa_1/\kappa_2=1$.}
  \label{fig:8}
\end{figure*}

Next, we investigate the influence of mechanical damping $\gamma$ on transmission properties. Figures \ref{fig:7} and \ref{fig:8} present transmission amplitudes $\mathcal{T}_{12}$ and $\mathcal{T}_{21}$ as functions of detuning for different values of the mechanical damping rate $\gamma$. The mechanical damping rate $\gamma$ dictates the rate at which energy is dissipated in the mechanical oscillator. Upon varying $\gamma$, we observe the interesting behaviors in the transmission amplitudes. In fact, for low mechanical damping rates ($\gamma/\kappa_2 = 0.001$, Figures \ref{fig:7}(a) and \ref{fig:8}(a)), the transmission amplitudes exhibit sharp peaks around the resonance. The detuning range conducive to nonreciprocal transmission is narrow, making the system sensitive to minor variations in detuning. In this regime, mechanical damping is insufficient to substantially influence nonreciprocity. As the mechanical damping rate increases to $\gamma/\kappa_2 = 0.01$ (Figures \ref{fig:7}(b) and \ref{fig:8}(b)), there is a broadening of the nonreciprocal transmission range. The peaks of $\mathcal{T}_{12}$ become wider and $\mathcal{T}_{21}$ decreases, indicating enhanced nonreciprocity. Mechanical damping assumes a more substantial role in modulating the system's behavior, effectively suppressing back-propagating signals. At higher mechanical damping rates ($\gamma/\kappa_2 = 0.1$ and $\gamma/\kappa_2 = 1$, Figures \ref{fig:7}(c-d) and \ref{fig:8}(c-d)), the detuning range over which nonreciprocity occurs is significantly broadened. The peaks of $\mathcal{T}_{12}$ and $\mathcal{T}_{21}$ flatten, suggesting that the system may maintain nonreciprocity over an extensive detuning range. In this regime, mechanical damping exerts sufficient influence to dominate the system's dynamics, culminating in robust nonreciprocal transmission that is less susceptible to detuning variations.

The findings of Figures \ref{fig:7} and \ref{fig:8} illustrate that an increase in mechanical damping enhances the robustness of nonreciprocity by expanding the de-tuning range for nonreciprocal transmission. This suggests that mechanical dissipation serves as a stabilizing function, akin to the decay of the atomic ensemble, in the maintenance of nonreciprocal transmission.  Overall, the results presented above illustrate that, through modulation of phase differences, coupling strengths, and dissipation rates, it is feasible to achieve and regulate perfect nonreciprocity in the hybrid optomechanical system. 
\begin{figure}[htp!]
	\centering
	{\includegraphics[width=9cm]{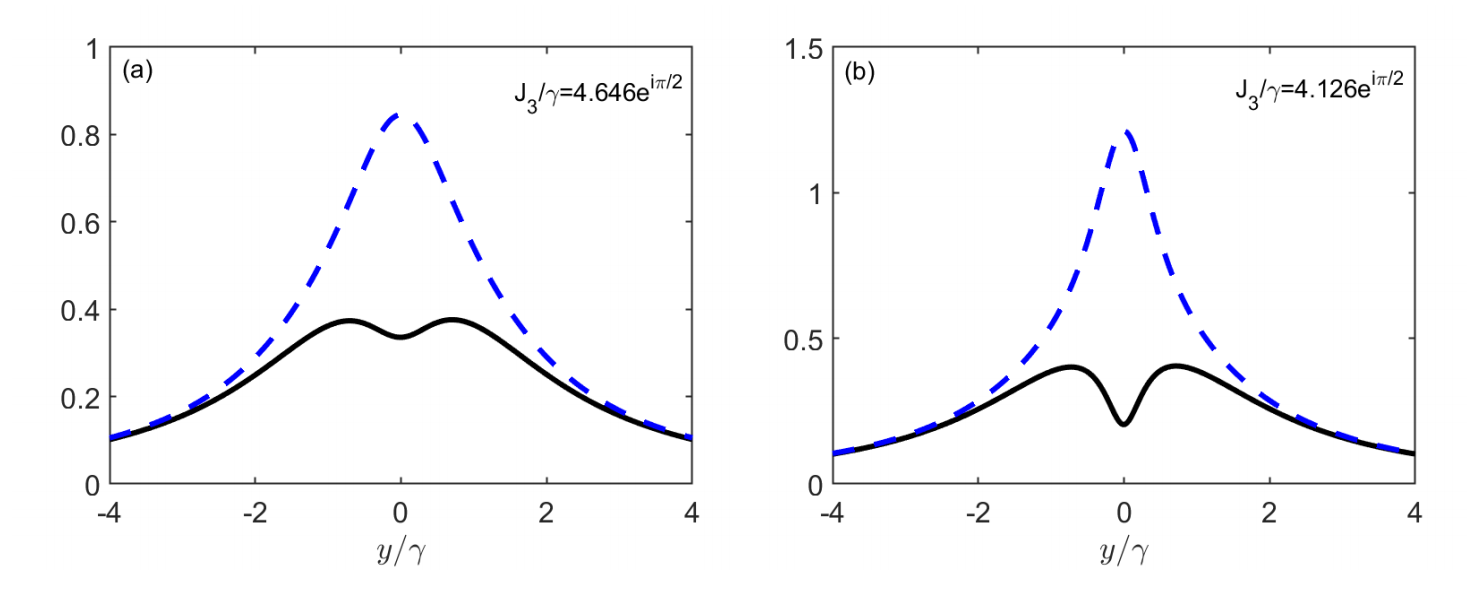}}
	\caption{Transmission amplitudes $\mathcal{T}_{12}$ (black solid--line) and $\mathcal{T}_{21}$ (blue dotted--line) as a function of detuning $y/\gamma$ for different values of $J_3$. Parameters used are the same as for Figure \ref{fig:2} except for  $\theta=\phi=\frac{\pi}{2}$ and complex coupling $J_3$ (a) $J_3/\gamma=4.646e^{i\frac{\pi}{2}}$, (b) $J_3/\gamma=4.126e^{i\frac{\pi}{2}}$ .}
	\label{fig:9}	
\end{figure}
These outcomes provide a foundation for the design of nonreciprocal optical devices with implications in photonics and quantum information processing.

It can be clearly seen from \Cref{fig:9}, that our system can show gain (amplification) and loss (decay) by carefully adjusting the value of the complex coupling constant $J_3$.

\section{Conclusion} \label{sec:IV}

This paper presents a detailed analysis of a hybrid optomechanical system that incorporates an atomic ensemble to achieve perfect optical nonreciprocal transmission. By introducing complex coupling strengths between the atomic ensemble and the mechanical oscillator, it is demonstrated that nonreciprocal transmission results from interference among distinct optical paths within the system. Specifically, the study reveals that the real and imaginary components of the coupling constants, along with the relative phase differences among the optomechanical coupling strengths, are essential in determining the directionality of optical signals. Through meticulous tuning of system parameters—such as coupling strengths, detuning, and phase differences—optimal nonreciprocal transmission was achieved under specific resonance conditions. Furthermore, the conditions required for perfect nonreciprocity were derived, and explicit expressions for the critical system parameters were provided. These findings offer a foundational framework for the design of optical isolators and circulators, which are essential components in quantum communication and signal processing applications. 

Our analysis relies on an effective non-Hermitian Hamiltonian to model the dissipative coupling between the atomic ensemble and the mechanical oscillator, which simplifies the treatment of environmental interactions. Although this approach successfully captures the qualitative features of nonreciprocal transmission, it is an approximation that does not fully describe microscopic bath dynamics or stochastic quantum jumps, which would be taken into account in a complete Lindblad master equation framework. Future research may explore the experimental realization of this system, with a particular focus on the role of dissipative couplings and environmental interactions in maintaining nonreciprocity. Moreover, investigating the robustness of nonreciprocal transmission in the presence of system imperfections and noise would yield significant insights. Overall, the results advance the development of sophisticated nonreciprocal optical devices with prospective applications in quantum information processing, photonics, and precision metrology.

\section*{Acknowledgments}

This work has been carried out under the Iso-Lomso Fellowship at Stellenbosch Institute for Advanced Study (STIAS), Wallenberg Research Centre at Stellenbosch University, Stellenbosch 7600, South Africa. K.S. Nisar is grateful to the funding from Prince Sattam bin Abdulaziz University, Saudi Arabia project number (PSAU/2024/R/1445). The authors are thankful to the Deanship of Graduate Studies and Scientific Research at University of Bisha for supporting this work through the Fast-Track Research Support Program.

 \textbf{Author Contributions:} E.K.B. and J.-X. conceptualized the work and carry out the simulations and analysis. A., P.D., and S.G.N.E.  participated in all the discussions and provided useful suggestions to the final version of the manuscript. K.S.N. and A.-H.A.-A. supervised the work. All authors participated equally in the discussions and the preparation of the final manuscript.

\textbf{Competing Interests:} All authors declare no competing interests.

 \section*{Data Availability}
Relevant data are included in the manuscript and supporting information. Supplement data are available upon reasonable request.

\section*{Appendix: Expressions of $R_7, R_8$, and $R_9$}

 The determinant of $A_1$ is given by the expression
\begin{equation}
\begin{aligned}
\mathcal{D}&=-2J_1|J_2|J_3G_2\cos(\theta-\phi)-2i|J_2|J_3\kappa_2G_1\cos\phi\\&-2iJ_1G_1G_2f\cos\theta-2|J_2|J_3G_1y\cos\phi-2G_1G_2y\cos\theta\\&+G^2_2\mathcal{D}_1+G^2_1\mathcal{D}_2+|J_2|^2\mathcal{D}_3+J^2_1\mathcal{D}_{4}+J^2_3\mathcal{D}_{5}+y^3\mathcal{D}_{6}\\&+y^2\mathcal{D}_{7}+\kappa_1\kappa_2\mathcal{D}_{8}-i\gamma fy\mathcal{D}_{9}+y^4,
\end{aligned}
\end{equation}
where 
\begin{equation}
\begin{aligned}
\mathcal{D}_1&=|J_2|^2-iy\kappa_1-ify+f\kappa_1-y^2,\\
\mathcal{D}_2&=-y^2-ify-iy\kappa_2+f\kappa_2,\\
\mathcal{D}_3&=-i\gamma y-iy\kappa_2+\gamma\kappa_2-y^2,\\
\mathcal{D}_{4}&=-y^2+J^2_3-i\gamma y-ify+\gamma f,\\
\mathcal{D}_{5}&=-y^2-iy\kappa_1-iy\kappa_2+\kappa_1\kappa_2,\\
\mathcal{D}_{6}&=i\kappa_1+i\kappa_2+i\gamma+if,\\
\mathcal{D}_{7}&=-\gamma f-\gamma\kappa_2-f\kappa_1-\kappa_1\kappa_2-f\kappa_2-\gamma\kappa_1,\\
\mathcal{D}_{8}&=\gamma f-ify-i\gamma y,~~\mathcal{D}_{9}=\kappa_1+\kappa_2.
\end{aligned}
\end{equation}
For simplicity, we consider the case where $\theta=\phi=\frac{\pi}{2}$, and the expression of $\mathcal{D}$ becomes
$\mathcal{D}=J_3^2(R_5+R_2^\prime)+R_4+\frac{R_6}{J_3^2}(J_1J_3^2+J_1\gamma f-G_1G_2f)^2+R_3$, where $R_1=\frac{1}{2\sqrt{\kappa_1\kappa_2}}$, $R_2=\frac{-2G_1G_2f}{R_1}$, $R^\prime_2=J_1^2+\kappa_1\kappa_2$, $R_3=G^2_2f\kappa_1+G_1^2f\kappa_2+J_1^2\gamma f+\kappa_1\kappa_2\gamma f$, $R_4=-2J_1f(J_1\gamma-G_1G_2)$, $R_5=-2J_1^2$, $R_6=\frac{G_2^2+\gamma\kappa_2}{G_2^2}$, $R_7=R_5+R_2^\prime+R_6J_1^2$, $R_8=R_4-R_2+2R_6J_1^2\gamma f-2R_6J_1G_1G_2 f+R_3$, and $R_9=R_6J_1^2\gamma^2f^2-2R_6J_1\gamma f^2G_1G_2+R_6G_1^2G_2^2f^2$.

\bibliography{Refn}

\begin{thebibliography}{27}%
\makeatletter
\providecommand \@ifxundefined [1]{%
 \@ifx{#1\undefined}
}%
\providecommand \@ifnum [1]{%
 \ifnum #1\expandafter \@firstoftwo
 \else \expandafter \@secondoftwo
 \fi
}%
\providecommand \@ifx [1]{%
 \ifx #1\expandafter \@firstoftwo
 \else \expandafter \@secondoftwo
 \fi
}%
\providecommand \natexlab [1]{#1}%
\providecommand \enquote  [1]{``#1''}%
\providecommand \bibnamefont  [1]{#1}%
\providecommand \bibfnamefont [1]{#1}%
\providecommand \citenamefont [1]{#1}%
\providecommand \href@noop [0]{\@secondoftwo}%
\providecommand \href [0]{\begingroup \@sanitize@url \@href}%
\providecommand \@href[1]{\@@startlink{#1}\@@href}%
\providecommand \@@href[1]{\endgroup#1\@@endlink}%
\providecommand \@sanitize@url [0]{\catcode `\\12\catcode `\$12\catcode
  `\&12\catcode `\#12\catcode `\^12\catcode `\_12\catcode `\%12\relax}%
\providecommand \@@startlink[1]{}%
\providecommand \@@endlink[0]{}%
\providecommand \url  [0]{\begingroup\@sanitize@url \@url }%
\providecommand \@url [1]{\endgroup\@href {#1}{\urlprefix }}%
\providecommand \urlprefix  [0]{URL }%
\providecommand \Eprint [0]{\href }%
\providecommand \doibase [0]{https://doi.org/}%
\providecommand \selectlanguage [0]{\@gobble}%
\providecommand \bibinfo  [0]{\@secondoftwo}%
\providecommand \bibfield  [0]{\@secondoftwo}%
\providecommand \translation [1]{[#1]}%
\providecommand \BibitemOpen [0]{}%
\providecommand \bibitemStop [0]{}%
\providecommand \bibitemNoStop [0]{.\EOS\space}%
\providecommand \EOS [0]{\spacefactor3000\relax}%
\providecommand \BibitemShut  [1]{\csname bibitem#1\endcsname}%
\let\auto@bib@innerbib\@empty
\bibitem [{\citenamefont {Meiser}\ and\ \citenamefont
  {Meystre}(2006)}]{Meiser2006}%
  \BibitemOpen
  \bibfield  {author} {\bibinfo {author} {\bibfnamefont {D.}~\bibnamefont
  {Meiser}}\ and\ \bibinfo {author} {\bibfnamefont {P.}~\bibnamefont
  {Meystre}},\ }\bibfield  {title} {\bibinfo {title} {Coupled dynamics of atoms
  and radiation-pressure-driven interferometers},\ }\href
  {https://doi.org/10.1103/PhysRevA.73.033417} {\bibfield  {journal} {\bibinfo
  {journal} {Phys. Rev. A}\ }\textbf {\bibinfo {volume} {73}},\ \bibinfo
  {pages} {033417} (\bibinfo {year} {2006})}\BibitemShut {NoStop}%
\bibitem [{\citenamefont {Agasti}\ and\ \citenamefont
  {Djorwé}(2024)}]{Agasti2024}%
  \BibitemOpen
  \bibfield  {author} {\bibinfo {author} {\bibfnamefont {S.}~\bibnamefont
  {Agasti}}\ and\ \bibinfo {author} {\bibfnamefont {P.}~\bibnamefont
  {Djorwé}},\ }\bibfield  {title} {\bibinfo {title} {Bistability-assisted
  mechanical squeezing and entanglement},\ }\href
  {https://doi.org/10.1088/1402-4896/ad6eca} {\bibfield  {journal} {\bibinfo
  {journal} {Physica Scripta}\ }\textbf {\bibinfo {volume} {99}},\ \bibinfo
  {pages} {095129} (\bibinfo {year} {2024})}\BibitemShut {NoStop}%
\bibitem [{\citenamefont {Bhattacherjee}(2009)}]{Bhattacherjee2009}%
  \BibitemOpen
  \bibfield  {author} {\bibinfo {author} {\bibfnamefont {A.~B.}\ \bibnamefont
  {Bhattacherjee}},\ }\bibfield  {title} {\bibinfo {title} {Cavity quantum
  optomechanics of ultracold atoms in an optical lattice: Normal-mode
  splitting},\ }\href {https://doi.org/10.1103/PhysRevA.80.043607} {\bibfield
  {journal} {\bibinfo  {journal} {Phys. Rev. A}\ }\textbf {\bibinfo {volume}
  {80}},\ \bibinfo {pages} {043607} (\bibinfo {year} {2009})}\BibitemShut
  {NoStop}%
\bibitem [{\citenamefont {Paternostro}\ \emph {et~al.}(2010)\citenamefont
  {Paternostro}, \citenamefont {De~Chiara},\ and\ \citenamefont
  {Palma}}]{Paternostro2010}%
  \BibitemOpen
  \bibfield  {author} {\bibinfo {author} {\bibfnamefont {M.}~\bibnamefont
  {Paternostro}}, \bibinfo {author} {\bibfnamefont {G.}~\bibnamefont
  {De~Chiara}},\ and\ \bibinfo {author} {\bibfnamefont {G.~M.}\ \bibnamefont
  {Palma}},\ }\bibfield  {title} {\bibinfo {title} {Cold-atom-induced control
  of an optomechanical device},\ }\href
  {https://doi.org/10.1103/PhysRevLett.104.243602} {\bibfield  {journal}
  {\bibinfo  {journal} {Phys. Rev. Lett.}\ }\textbf {\bibinfo {volume} {104}},\
  \bibinfo {pages} {243602} (\bibinfo {year} {2010})}\BibitemShut {NoStop}%
\bibitem [{\citenamefont {Jin}\ \emph {et~al.}(2021)\citenamefont {Jin},
  \citenamefont {Peng}, \citenamefont {Yuan},\ and\ \citenamefont
  {Feng}}]{jin2021macroscopic}%
  \BibitemOpen
  \bibfield  {author} {\bibinfo {author} {\bibfnamefont {L.}~\bibnamefont
  {Jin}}, \bibinfo {author} {\bibfnamefont {J.-X.}\ \bibnamefont {Peng}},
  \bibinfo {author} {\bibfnamefont {Q.-Z.}\ \bibnamefont {Yuan}},\ and\
  \bibinfo {author} {\bibfnamefont {X.-L.}\ \bibnamefont {Feng}},\ }\bibfield
  {title} {\bibinfo {title} {Macroscopic quantum coherence in a spinning
  optomechanical system},\ }\href {https://doi.org/10.1364/oe.443486}
  {\bibfield  {journal} {\bibinfo  {journal} {Optics Express}\ }\textbf
  {\bibinfo {volume} {29}},\ \bibinfo {pages} {41191} (\bibinfo {year}
  {2021})}\BibitemShut {NoStop}%
\bibitem [{\citenamefont {Massembele}\ \emph {et~al.}(2024)\citenamefont
  {Massembele}, \citenamefont {Djorwé}, \citenamefont {Sarma}, \citenamefont
  {Abdel-Aty},\ and\ \citenamefont {Engo}}]{Rostand2024}%
  \BibitemOpen
  \bibfield  {author} {\bibinfo {author} {\bibfnamefont {D.~R.~K.}\
  \bibnamefont {Massembele}}, \bibinfo {author} {\bibfnamefont
  {P.}~\bibnamefont {Djorwé}}, \bibinfo {author} {\bibfnamefont {A.~K.}\
  \bibnamefont {Sarma}}, \bibinfo {author} {\bibfnamefont {A.-H.}\ \bibnamefont
  {Abdel-Aty}},\ and\ \bibinfo {author} {\bibfnamefont {S.~G.~N.}\ \bibnamefont
  {Engo}},\ }\bibfield  {title} {\bibinfo {title} {Quantum entanglement
  assisted via duffing nonlinearity},\ }\href
  {https://doi.org/10.1103/PhysRevA.110.043502} {\bibfield  {journal} {\bibinfo
   {journal} {Physical Review A}\ }\textbf {\bibinfo {volume} {110}},\ \bibinfo
  {pages} {043502} (\bibinfo {year} {2024})}\BibitemShut {NoStop}%
\bibitem [{\citenamefont {Blais}\ \emph {et~al.}(2020)\citenamefont {Blais},
  \citenamefont {Girvin},\ and\ \citenamefont {Oliver}}]{Blais_2020}%
  \BibitemOpen
  \bibfield  {author} {\bibinfo {author} {\bibfnamefont {A.}~\bibnamefont
  {Blais}}, \bibinfo {author} {\bibfnamefont {S.~M.}\ \bibnamefont {Girvin}},\
  and\ \bibinfo {author} {\bibfnamefont {W.~D.}\ \bibnamefont {Oliver}},\
  }\bibfield  {title} {\bibinfo {title} {Quantum information processing and
  quantum optics with circuit quantum electrodynamics},\ }\href
  {https://doi.org/10.1038/s41567-020-0806-z} {\bibfield  {journal} {\bibinfo
  {journal} {Nature Physics}\ }\textbf {\bibinfo {volume} {16}},\ \bibinfo
  {pages} {247} (\bibinfo {year} {2020})}\BibitemShut {NoStop}%
\bibitem [{\citenamefont {Stannigel}\ \emph {et~al.}(2012)\citenamefont
  {Stannigel}, \citenamefont {Komar}, \citenamefont {Habraken}, \citenamefont
  {Bennett}, \citenamefont {Lukin}, \citenamefont {Zoller},\ and\ \citenamefont
  {Rabl}}]{Stannigel_2012}%
  \BibitemOpen
  \bibfield  {author} {\bibinfo {author} {\bibfnamefont {K.}~\bibnamefont
  {Stannigel}}, \bibinfo {author} {\bibfnamefont {P.}~\bibnamefont {Komar}},
  \bibinfo {author} {\bibfnamefont {S.~J.~M.}\ \bibnamefont {Habraken}},
  \bibinfo {author} {\bibfnamefont {S.~D.}\ \bibnamefont {Bennett}}, \bibinfo
  {author} {\bibfnamefont {M.~D.}\ \bibnamefont {Lukin}}, \bibinfo {author}
  {\bibfnamefont {P.}~\bibnamefont {Zoller}},\ and\ \bibinfo {author}
  {\bibfnamefont {P.}~\bibnamefont {Rabl}},\ }\bibfield  {title} {\bibinfo
  {title} {Optomechanical quantum information processing with photons and
  phonons},\ }\href {https://doi.org/10.1103/PhysRevLett.109.013603} {\bibfield
   {journal} {\bibinfo  {journal} {Physical Review Letters}\ }\textbf {\bibinfo
  {volume} {109}},\ \bibinfo {pages} {013603} (\bibinfo {year}
  {2012})}\BibitemShut {NoStop}%
\bibitem [{\citenamefont {Djorwe}\ \emph {et~al.}(2019)\citenamefont {Djorwe},
  \citenamefont {Pennec},\ and\ \citenamefont {Djafari-Rouhani}}]{Djorwe_2019}%
  \BibitemOpen
  \bibfield  {author} {\bibinfo {author} {\bibfnamefont {P.}~\bibnamefont
  {Djorwe}}, \bibinfo {author} {\bibfnamefont {Y.}~\bibnamefont {Pennec}},\
  and\ \bibinfo {author} {\bibfnamefont {B.}~\bibnamefont {Djafari-Rouhani}},\
  }\bibfield  {title} {\bibinfo {title} {Exceptional point enhances sensitivity
  of optomechanical mass sensors},\ }\bibfield  {journal} {\bibinfo  {journal}
  {Physical Review Applied}\ }\textbf {\bibinfo {volume} {12}},\ \href
  {https://doi.org/10.1103/PhysRevApplied.12.024002}
  {10.1103/PhysRevApplied.12.024002} (\bibinfo {year} {2019})\BibitemShut
  {NoStop}%
\bibitem [{\citenamefont {Tchounda}\ \emph {et~al.}(2023)\citenamefont
  {Tchounda}, \citenamefont {Djorwé}, \citenamefont {Engo},\ and\
  \citenamefont {Djafari-Rouhani}}]{Tchounda_2023}%
  \BibitemOpen
  \bibfield  {author} {\bibinfo {author} {\bibfnamefont {S.~M.}\ \bibnamefont
  {Tchounda}}, \bibinfo {author} {\bibfnamefont {P.}~\bibnamefont {Djorwé}},
  \bibinfo {author} {\bibfnamefont {S.~N.}\ \bibnamefont {Engo}},\ and\
  \bibinfo {author} {\bibfnamefont {B.}~\bibnamefont {Djafari-Rouhani}},\
  }\bibfield  {title} {\bibinfo {title} {Sensor sensitivity based on
  exceptional points engineered via synthetic magnetism},\ }\href
  {https://doi.org/10.1103/PhysRevApplied.19.064016} {\bibfield  {journal}
  {\bibinfo  {journal} {Physical Review Applied}\ }\textbf {\bibinfo {volume}
  {19}},\ \bibinfo {pages} {064016} (\bibinfo {year} {2023})}\BibitemShut
  {NoStop}%
\bibitem [{\citenamefont {Djorwé}\ \emph {et~al.}(2024)\citenamefont
  {Djorwé}, \citenamefont {Asjad}, \citenamefont {Pennec}, \citenamefont
  {Dutykh},\ and\ \citenamefont {Djafari-Rouhani}}]{Djor2024}%
  \BibitemOpen
  \bibfield  {author} {\bibinfo {author} {\bibfnamefont {P.}~\bibnamefont
  {Djorwé}}, \bibinfo {author} {\bibfnamefont {M.}~\bibnamefont {Asjad}},
  \bibinfo {author} {\bibfnamefont {Y.}~\bibnamefont {Pennec}}, \bibinfo
  {author} {\bibfnamefont {D.}~\bibnamefont {Dutykh}},\ and\ \bibinfo {author}
  {\bibfnamefont {B.}~\bibnamefont {Djafari-Rouhani}},\ }\bibfield  {title}
  {\bibinfo {title} {Parametrically enhancing sensor sensitivity at an
  exceptional point},\ }\href
  {https://doi.org/10.1103/PhysRevResearch.6.033284} {\bibfield  {journal}
  {\bibinfo  {journal} {Physical Review Research}\ }\textbf {\bibinfo {volume}
  {6}},\ \bibinfo {pages} {033284} (\bibinfo {year} {2024})}\BibitemShut
  {NoStop}%
\bibitem [{\citenamefont {Metcalfe}(2014)}]{Metcalfe:2014jnz}%
  \BibitemOpen
  \bibfield  {author} {\bibinfo {author} {\bibfnamefont {M.}~\bibnamefont
  {Metcalfe}},\ }\bibfield  {title} {\bibinfo {title} {{Applications of cavity
  optomechanics}},\ }\href {https://doi.org/10.1063/1.4896029} {\bibfield
  {journal} {\bibinfo  {journal} {Appl. Phys. Rev.}\ }\textbf {\bibinfo
  {volume} {1}},\ \bibinfo {pages} {031105} (\bibinfo {year}
  {2014})}\BibitemShut {NoStop}%
\bibitem [{\citenamefont {Aspelmeyer}\ \emph {et~al.}(2014)\citenamefont
  {Aspelmeyer}, \citenamefont {Kippenberg},\ and\ \citenamefont
  {Marquardt}}]{Aspelmeyer2014}%
  \BibitemOpen
  \bibfield  {author} {\bibinfo {author} {\bibfnamefont {M.}~\bibnamefont
  {Aspelmeyer}}, \bibinfo {author} {\bibfnamefont {T.~J.}\ \bibnamefont
  {Kippenberg}},\ and\ \bibinfo {author} {\bibfnamefont {F.}~\bibnamefont
  {Marquardt}},\ }\bibfield  {title} {\bibinfo {title} {Cavity optomechanics},\
  }\href {https://doi.org/10.1103/RevModPhys.86.1391} {\bibfield  {journal}
  {\bibinfo  {journal} {Rev. Mod. Phys.}\ }\textbf {\bibinfo {volume} {86}},\
  \bibinfo {pages} {1391} (\bibinfo {year} {2014})}\BibitemShut {NoStop}%
\bibitem [{\citenamefont {Xu}\ \emph {et~al.}(2024)\citenamefont {Xu},
  \citenamefont {Zhu}, \citenamefont {Chen}, \citenamefont {Li},\ and\
  \citenamefont {Zhang}}]{Xu_2024}%
  \BibitemOpen
  \bibfield  {author} {\bibinfo {author} {\bibfnamefont {X.}~\bibnamefont
  {Xu}}, \bibinfo {author} {\bibfnamefont {H.}~\bibnamefont {Zhu}}, \bibinfo
  {author} {\bibfnamefont {S.}~\bibnamefont {Chen}}, \bibinfo {author}
  {\bibfnamefont {F.}~\bibnamefont {Li}},\ and\ \bibinfo {author}
  {\bibfnamefont {X.}~\bibnamefont {Zhang}},\ }\bibfield  {title} {\bibinfo
  {title} {Nonlinear dynamics of cavity optomechanical-thermal systems},\
  }\href {https://doi.org/10.1364/OE.515095} {\bibfield  {journal} {\bibinfo
  {journal} {Optics Express}\ }\textbf {\bibinfo {volume} {32}},\ \bibinfo
  {pages} {7611} (\bibinfo {year} {2024})}\BibitemShut {NoStop}%
\bibitem [{\citenamefont {Djorwe}\ \emph {et~al.}(2022)\citenamefont {Djorwe},
  \citenamefont {Yves Effa},\ and\ \citenamefont {G. Nana Engo}}]{Djor2022}%
  \BibitemOpen
  \bibfield  {author} {\bibinfo {author} {\bibfnamefont {P.}~\bibnamefont
  {Djorwe}}, \bibinfo {author} {\bibfnamefont {J.}~\bibnamefont {Yves Effa}},\
  and\ \bibinfo {author} {\bibfnamefont {S.}~\bibnamefont {G. Nana Engo}},\
  }\bibfield  {title} {\bibinfo {title} {Hidden attractors and metamorphoses of
  basin boundaries in optomechanics},\ }\href
  {https://doi.org/10.1007/s11071-022-08139-2} {\bibfield  {journal} {\bibinfo
  {journal} {Nonlinear Dynamics}\ }\textbf {\bibinfo {volume} {111}},\ \bibinfo
  {pages} {5905} (\bibinfo {year} {2022})}\BibitemShut {NoStop}%
\bibitem [{\citenamefont {Koch}\ \emph {et~al.}(2010)\citenamefont {Koch},
  \citenamefont {Houck}, \citenamefont {Hur},\ and\ \citenamefont
  {Girvin}}]{Koch2010}%
  \BibitemOpen
  \bibfield  {author} {\bibinfo {author} {\bibfnamefont {J.}~\bibnamefont
  {Koch}}, \bibinfo {author} {\bibfnamefont {A.~A.}\ \bibnamefont {Houck}},
  \bibinfo {author} {\bibfnamefont {K.~L.}\ \bibnamefont {Hur}},\ and\ \bibinfo
  {author} {\bibfnamefont {S.~M.}\ \bibnamefont {Girvin}},\ }\bibfield  {title}
  {\bibinfo {title} {Time-reversal-symmetry breaking in circuit-qed-based
  photon lattices},\ }\href {https://doi.org/10.1103/PhysRevA.82.043811}
  {\bibfield  {journal} {\bibinfo  {journal} {Phys. Rev. A}\ }\textbf {\bibinfo
  {volume} {82}},\ \bibinfo {pages} {043811} (\bibinfo {year}
  {2010})}\BibitemShut {NoStop}%
\bibitem [{\citenamefont {Metelmann}\ and\ \citenamefont
  {Clerk}(2015)}]{Metelmann2015}%
  \BibitemOpen
  \bibfield  {author} {\bibinfo {author} {\bibfnamefont {A.}~\bibnamefont
  {Metelmann}}\ and\ \bibinfo {author} {\bibfnamefont {A.}~\bibnamefont
  {Clerk}},\ }\bibfield  {title} {\bibinfo {title} {Nonreciprocal photon
  transmission and amplification via reservoir engineering},\ }\href
  {https://doi.org/10.1103/physrevx.5.021025} {\bibfield  {journal} {\bibinfo
  {journal} {Physical Review X}\ }\textbf {\bibinfo {volume} {5}},\ \bibinfo
  {pages} {021025} (\bibinfo {year} {2015})}\BibitemShut {NoStop}%
\bibitem [{\citenamefont {Solja\v{c}i\'{c}}\ \emph {et~al.}(2003)\citenamefont
  {Solja\v{c}i\'{c}}, \citenamefont {Luo}, \citenamefont {Joannopoulos},\ and\
  \citenamefont {Fan}}]{Soljacic:03}%
  \BibitemOpen
  \bibfield  {author} {\bibinfo {author} {\bibfnamefont {M.}~\bibnamefont
  {Solja\v{c}i\'{c}}}, \bibinfo {author} {\bibfnamefont {C.}~\bibnamefont
  {Luo}}, \bibinfo {author} {\bibfnamefont {J.~D.}\ \bibnamefont
  {Joannopoulos}},\ and\ \bibinfo {author} {\bibfnamefont {S.}~\bibnamefont
  {Fan}},\ }\bibfield  {title} {\bibinfo {title} {Nonlinear photonic crystal
  microdevices for optical integration},\ }\href
  {https://doi.org/10.1364/OL.28.000637} {\bibfield  {journal} {\bibinfo
  {journal} {Opt. Lett.}\ }\textbf {\bibinfo {volume} {28}},\ \bibinfo {pages}
  {637} (\bibinfo {year} {2003})}\BibitemShut {NoStop}%
\bibitem [{\citenamefont {He}\ \emph {et~al.}(2018)\citenamefont {He},
  \citenamefont {Yang}, \citenamefont {Jiang},\ and\ \citenamefont
  {Xiao}}]{He2018}%
  \BibitemOpen
  \bibfield  {author} {\bibinfo {author} {\bibfnamefont {B.}~\bibnamefont
  {He}}, \bibinfo {author} {\bibfnamefont {L.}~\bibnamefont {Yang}}, \bibinfo
  {author} {\bibfnamefont {X.}~\bibnamefont {Jiang}},\ and\ \bibinfo {author}
  {\bibfnamefont {M.}~\bibnamefont {Xiao}},\ }\bibfield  {title} {\bibinfo
  {title} {Transmission nonreciprocity in a mutually coupled circulating
  structure},\ }\href {https://doi.org/10.1103/PhysRevLett.120.203904}
  {\bibfield  {journal} {\bibinfo  {journal} {Phys. Rev. Lett.}\ }\textbf
  {\bibinfo {volume} {120}},\ \bibinfo {pages} {203904} (\bibinfo {year}
  {2018})}\BibitemShut {NoStop}%
\bibitem [{\citenamefont {El-Ganainy}\ \emph {et~al.}(2018)\citenamefont
  {El-Ganainy}, \citenamefont {Makris}, \citenamefont {Khajavikhan},
  \citenamefont {Musslimani}, \citenamefont {Rotter},\ and\ \citenamefont
  {Christodoulides}}]{el2018non}%
  \BibitemOpen
  \bibfield  {author} {\bibinfo {author} {\bibfnamefont {R.}~\bibnamefont
  {El-Ganainy}}, \bibinfo {author} {\bibfnamefont {K.~G.}\ \bibnamefont
  {Makris}}, \bibinfo {author} {\bibfnamefont {M.}~\bibnamefont {Khajavikhan}},
  \bibinfo {author} {\bibfnamefont {Z.~H.}\ \bibnamefont {Musslimani}},
  \bibinfo {author} {\bibfnamefont {S.}~\bibnamefont {Rotter}},\ and\ \bibinfo
  {author} {\bibfnamefont {D.~N.}\ \bibnamefont {Christodoulides}},\ }\bibfield
   {title} {\bibinfo {title} {Non-hermitian physics and pt symmetry},\ }\href
  {https://doi.org/10.1038/nphys4323} {\bibfield  {journal} {\bibinfo
  {journal} {Nature Physics}\ }\textbf {\bibinfo {volume} {14}},\ \bibinfo
  {pages} {11} (\bibinfo {year} {2018})}\BibitemShut {NoStop}%
\bibitem [{\citenamefont {Ozdemir}\ \emph {et~al.}(2019)\citenamefont
  {Ozdemir}, \citenamefont {Rotter}, \citenamefont {Nori},\ and\ \citenamefont
  {Yang}}]{Oezdemir2019}%
  \BibitemOpen
  \bibfield  {author} {\bibinfo {author} {\bibfnamefont {S.~K.}\ \bibnamefont
  {Ozdemir}}, \bibinfo {author} {\bibfnamefont {S.}~\bibnamefont {Rotter}},
  \bibinfo {author} {\bibfnamefont {F.}~\bibnamefont {Nori}},\ and\ \bibinfo
  {author} {\bibfnamefont {L.}~\bibnamefont {Yang}},\ }\bibfield  {title}
  {\bibinfo {title} {Parity–time symmetry and exceptional points in
  photonics},\ }\href {https://doi.org/10.1038/s41563-019-0304-9} {\bibfield
  {journal} {\bibinfo  {journal} {Nature Materials}\ }\textbf {\bibinfo
  {volume} {18}},\ \bibinfo {pages} {783} (\bibinfo {year} {2019})}\BibitemShut
  {NoStop}%
\bibitem [{\citenamefont {Mbokop~Tchounda}\ \emph {et~al.}(2024)\citenamefont
  {Mbokop~Tchounda}, \citenamefont {Djorwé}, \citenamefont {Tchakui},\ and\
  \citenamefont {Nana~Engo}}]{Mbokop2024}%
  \BibitemOpen
  \bibfield  {author} {\bibinfo {author} {\bibfnamefont {S.~R.}\ \bibnamefont
  {Mbokop~Tchounda}}, \bibinfo {author} {\bibfnamefont {P.}~\bibnamefont
  {Djorwé}}, \bibinfo {author} {\bibfnamefont {M.~V.}\ \bibnamefont
  {Tchakui}},\ and\ \bibinfo {author} {\bibfnamefont {S.~G.}\ \bibnamefont
  {Nana~Engo}},\ }\bibfield  {title} {\bibinfo {title} {Chaos control and
  exceptional point engineering via dissipative optomechanical coupling},\
  }\href {https://doi.org/10.1088/1402-4896/ad195c} {\bibfield  {journal}
  {\bibinfo  {journal} {Physica Scripta}\ }\textbf {\bibinfo {volume} {99}},\
  \bibinfo {pages} {025215} (\bibinfo {year} {2024})}\BibitemShut {NoStop}%
\bibitem [{\citenamefont {Hafezi}\ and\ \citenamefont {Rabl}(2012)}]{article8}%
  \BibitemOpen
  \bibfield  {author} {\bibinfo {author} {\bibfnamefont {M.}~\bibnamefont
  {Hafezi}}\ and\ \bibinfo {author} {\bibfnamefont {P.}~\bibnamefont {Rabl}},\
  }\bibfield  {title} {\bibinfo {title} {Optomechanically induced
  non-reciprocity in microring resonators},\ }\href
  {https://doi.org/10.1364/OE.20.007672} {\bibfield  {journal} {\bibinfo
  {journal} {Optics Express}\ }\textbf {\bibinfo {volume} {20}},\ \bibinfo
  {pages} {7672} (\bibinfo {year} {2012})}\BibitemShut {NoStop}%
\bibitem [{\citenamefont {Rotter}(2009)}]{rotter2009non}%
  \BibitemOpen
  \bibfield  {author} {\bibinfo {author} {\bibfnamefont {I.}~\bibnamefont
  {Rotter}},\ }\bibfield  {title} {\bibinfo {title} {A non-hermitian hamilton
  operator and the physics of open quantum systems},\ }\href
  {https://doi.org/10.1088/1751-8113/42/15/153001} {\bibfield  {journal}
  {\bibinfo  {journal} {Journal of Physics A: Mathematical and Theoretical}\
  }\textbf {\bibinfo {volume} {42}},\ \bibinfo {pages} {153001} (\bibinfo
  {year} {2009})}\BibitemShut {NoStop}%
\bibitem [{\citenamefont {Ashida}\ \emph {et~al.}(2020)\citenamefont {Ashida},
  \citenamefont {Gong},\ and\ \citenamefont {Ueda}}]{ashida2020non}%
  \BibitemOpen
  \bibfield  {author} {\bibinfo {author} {\bibfnamefont {Y.}~\bibnamefont
  {Ashida}}, \bibinfo {author} {\bibfnamefont {Z.}~\bibnamefont {Gong}},\ and\
  \bibinfo {author} {\bibfnamefont {M.}~\bibnamefont {Ueda}},\ }\bibfield
  {title} {\bibinfo {title} {Non-hermitian physics},\ }\href
  {https://doi.org/10.1080/00018732.2021.1876991} {\bibfield  {journal}
  {\bibinfo  {journal} {Advances in Physics}\ }\textbf {\bibinfo {volume}
  {69}},\ \bibinfo {pages} {249} (\bibinfo {year} {2020})}\BibitemShut
  {NoStop}%
\bibitem [{\citenamefont {Takata}\ \emph {et~al.}(2022)\citenamefont {Takata},
  \citenamefont {Roberts}, \citenamefont {Shinya},\ and\ \citenamefont
  {Notomi}}]{Takata2022}%
  \BibitemOpen
  \bibfield  {author} {\bibinfo {author} {\bibfnamefont {K.}~\bibnamefont
  {Takata}}, \bibinfo {author} {\bibfnamefont {N.}~\bibnamefont {Roberts}},
  \bibinfo {author} {\bibfnamefont {A.}~\bibnamefont {Shinya}},\ and\ \bibinfo
  {author} {\bibfnamefont {M.}~\bibnamefont {Notomi}},\ }\bibfield  {title}
  {\bibinfo {title} {Imaginary couplings in non-hermitian coupled-mode theory:
  Effects on exceptional points of optical resonators},\ }\href
  {https://doi.org/10.1103/PhysRevA.105.013523} {\bibfield  {journal} {\bibinfo
   {journal} {Phys. Rev. A}\ }\textbf {\bibinfo {volume} {105}},\ \bibinfo
  {pages} {013523} (\bibinfo {year} {2022})}\BibitemShut {NoStop}%
\bibitem [{\citenamefont {Xia}\ \emph {et~al.}(2019)\citenamefont {Xia},
  \citenamefont {Yan}, \citenamefont {Tian},\ and\ \citenamefont
  {Gao}}]{XIA2019197}%
  \BibitemOpen
  \bibfield  {author} {\bibinfo {author} {\bibfnamefont {C.-C.}\ \bibnamefont
  {Xia}}, \bibinfo {author} {\bibfnamefont {X.-B.}\ \bibnamefont {Yan}},
  \bibinfo {author} {\bibfnamefont {X.-D.}\ \bibnamefont {Tian}},\ and\
  \bibinfo {author} {\bibfnamefont {F.}~\bibnamefont {Gao}},\ }\bibfield
  {title} {\bibinfo {title} {Ideal optical isolator with a two-cavity
  optomechanical system},\ }\href
  {https://doi.org/https://doi.org/10.1016/j.optcom.2019.06.059} {\bibfield
  {journal} {\bibinfo  {journal} {Optics Communications}\ }\textbf {\bibinfo
  {volume} {451}},\ \bibinfo {pages} {197} (\bibinfo {year}
  {2019})}\BibitemShut {NoStop}%
\end{thebibliography}%

\end{document}